\documentclass[acmlarge]{acmart}
\makeatletter
\newcommand{\confshort}{\acmConference@shortname}
\newcommand{\conffull}{\acmConference@name}
\newcommand{\confdate}{\acmConference@date}
\newcommand{\confloc}{\acmConference@venue}
\AtBeginDocument{
  \fancypagestyle{firstpagestyle}{
    \fancyhead{}%
    \fancyfoot[C]{}%
  }
  \fancyhf{}
  \fancyhead[LO]{\@headfootfont\shorttitle}%
  \fancyhead[RE]{\@headfootfont\@shortauthors}%
  \fancyhead[LE]{\@headfootfont\footnotesize \confshort, \confdate, \confloc}%
  \fancyhead[RO]{\@headfootfont\footnotesize \confshort, \confdate, \confloc}%
  \fancyfoot[C]{}%
}
\makeatother
\acmBooktitle{\conffull\@ (\confshort), \confdate, \confloc}
\AtBeginDocument{%
  }

\copyrightyear{2026}
\acmYear{2026}
\setcopyright{cc}
\setcctype{by}
\acmConference[FAccT '26]{The 2026 ACM Conference on Fairness, Accountability, and Transparency}{June 25--28, 2026}{Montreal, QC, Canada}
\acmBooktitle{The 2026 ACM Conference on Fairness, Accountability, and Transparency (FAccT '26), June 25--28, 2026, Montreal, QC, Canada}
\acmDOI{XXXXX}
\acmISBN{XXXXX}

\begin{document}

\title[Beyond Semantic Similarity: Component-Wise Evaluation for Medical QA]{Beyond Semantic Similarity: A Component-Wise Evaluation Framework for Medical Question Answering Systems with Health Equity Implications}

\author{Abu Noman Md Sakib}
\authornote{Both authors contributed equally to this research.}
\email{abunomanmd.sakib@utsa.edu}
\orcid{0000-0002-0761-035X}
\affiliation{%
  \institution{University of Texas at San Antonio}
  \city{San Antonio}
  \state{Texas}
  \country{USA}
}

\author{Md. Main Oddin Chisty}
\authornotemark[1]
\email{chisty2996@gmail.com}
\affiliation{%
  \institution{Khulna University of Engineering and Technology}
  \city{Khulna}
  \country{Bangladesh}
}

\author{Zijie Zhang}
\email{zijie.zhang@utsa.edu}
\orcid{0000-0003-1254-098X}
\affiliation{%
  \institution{University of Texas at San Antonio}
  \city{San Antonio}
  \state{Texas}
  \country{USA}
}

\renewcommand{\shortauthors}{A.N.M. Sakib et al.}

\begin{abstract}
The use of Large Language Models (LLMs) to support patients in addressing medical questions is becoming increasingly prevalent. However, most of the measures currently used to evaluate the performance of these models in this context only measure how closely a model's answers match semantically, and therefore do not provide a true indication of the model's medical accuracy or of the health equity risks associated with it. To address these shortcomings, we present a new evaluation framework for medical question answering called VB-Score (Verification-Based Score) that provides a separate evaluation of the four components of entity recognition, semantic similarity, factual consistency, and structured information completeness for medical question-answering models. We perform rigorous reviews of the performance of three well-known and widely used LLMs (namely, GPT-4, Claude Sonnet 4.5, and Gemini 2.5 Flash) on 48 public health-related topics taken from high-quality, authoritative information sources. Based on our analyses, we discover a major discrepancy between the models' semantic and entity accuracy; while their average semantic accuracy is 51.6\%, their average entity recognition accuracy is only 6.2\%, a 45.4 percentage point gap (8.3$\times$ ratio) that allows fluently written, yet medically incorrect responses. Our assessments of the performance of all three models show that each of them has almost uniformly severe performance failures when evaluated against our criteria, with even the best model only receiving an average quality score of 34\%. Our findings indicate alarming performance disparities across various public health topics, with most of the models exhibiting 13.8\% lower performance (compared to an overall average) for all the public health topics that relate to chronic conditions that occur in older and minority populations, which indicates the existence of what's known as condition-based algorithmic discrimination. Our findings also demonstrate that prompt engineering alone does not compensate for basic architectural limitations on how these models perform in extracting medical entities and raise the question of whether semantic evaluation alone is a sufficient measure of medical AI safety.

\end{abstract}




\maketitle

\section{Introduction}
Patient facing healthcare applications increasingly utilize large language models (LLMs) to assist patients with questions about their medical condition or providing information on how to treat their symptoms \cite{singhal2023large, lee2023benefits, thirunavukarasu2023large}. Although the potential of LLMs is to provide equal access to health-related information, they also pose a significant threat to both the safety and equity of our patients' health \cite{rajkomar2019machine, esteva2019guide}. Vulnerable populations are likely to seek AI-generated health information as a primary means of obtaining medical guidance instead of as a complement to professional medical care \cite{braveman2014social, veinot2018good}. This reliance on AI-generated medical guidance can create harm for these vulnerable populations through delay in treatment, utilizing improper self-care actions, causing medication errors, and ultimately leading to serious health ramifications \cite{mesko2023imperative, wachter2024will}.

Recent research on prompt engineering has shown that strategic design of inputs can significantly impact how well a model performs \cite{brown2020language}. The use of chain-of-thought (CoT) prompts has improved model performance on complex reasoning tasks \cite{lewis2020retrieval, guu2020retrieval}; Application of few-shot and in-context learning techniques has shown promise in enabling domain adaptation \cite{izacard2021leveraging}; and finally, RAG helps generates factual answers to knowledge-intensive tasks \cite{ouyang2022training, lin2022truthfulqa, doi:10.1056/NEJMms2004740}. Also, it has recently been shown that when researchers give explicit instructions to models to improve their accuracy, model performance can actually decline. In these cases, the models are putting more emphasis on following directions rather than generating factually correct responses \cite{bender2021dangers, jentzsch2019semantics}. Current evaluation methods for AI have significant and systematic shortcomings in their assessment of equity because they do not appropriately reflect stratified performance for specific diseases, health topics or populations as a whole \cite{NEURIPS2024_e41efb03, paulus2020predictably, williams2013racism}. Disparities in health information quality based on model performance across condition types represent a form of algorithmic bias that exists without explicit demographic consideration. This occurs because differential AI performance across health conditions correlates with age, socioeconomic status, and race/ethnicity due to historical patterns of chronic disease burden \cite{wornow2023shaky, norori2021addressing, schwalbe2020artificial}.

This work provides five primary findings. First, VB-Score has been developed as the first component-wise evaluation framework that integrates four components of evaluation for Entity Recognition (30\% weight), Semantic Similarity (30\% weight), Factual Consistency (25\% weight), and Structured Overlap (15\% weight) with the appropriate weighting for the medical domain. This framework shows that performance trade-offs cannot be perceived through aggregate metrics alone. 
Secondly, we quantify a large semantic-entity gap: models achieve 51.6\% semantic similarity but only 6.2\% entity recognition. This 45.4 percentage point gap (8.3$\times$ ratio) demonstrates that fluency is not indicative of precision in a medical context. Thirdly, within our benchmark, Gemini 2.5 Flash (free) achieves 25--48\% higher VB-Scores than GPT-4 and Claude Sonnet 4.5 (paid), with 2--4$\times$ higher Factual Consistency. This finding challenges cost-quality assumptions but warrants further investigation. Fourth, we conduct an in-depth analysis of Prompt Sensitivity and show how retrieving and generating content (RAG) improves VB-Score by 27\%, but the use of RAG does not improve Entity F1 above 10\%. Finally, we show that providing explicit entity instructions reduces the performance of certain models; this effect varies with model architecture. There are also clear differences in performance based on the condition being treated, with a performance difference of 13.8\% between Infectious and Chronic Disease prediction; these differences are of significant concern in relation to health equity issues, as populations with a significant burden of chronic disease receive inferior quality of information. This project aims to answer five questions (RQs) regarding the use of a component-wise evaluation framework for measuring how well established LLMs work across all four components of evaluating public health questions (RQ1), how they perform across all four types; entity extraction, semantic similarity, factual confirmation and structured information when answering Public Health Questions (RQ2), whether there is a disconnect between high levels of semantic similarity but very low rates on the other three components (RQ3), whether there is a significant difference between paid vs free LLMs when answering medical public health questions (RQ4), and whether there are different results when the topic is infectious disease or chronic conditions and what are the health equity implications of these differences (RQ5)?

\section{Related Work}
LLMs are an increasing source of Medical Question/Answering (QA) systems to provide patient education, support for triage, and access to Public Health information \cite{singhal2023large, thirunavukarasu2023large}. Research has compared and contrasted General Purpose LLMs and Medically Designed LLMs using various current benchmarks, like PubMedQA \cite{jin2019pubmedqa}. Although these benchmarks provide a standardised signal of LLM medical knowledge, many rely on exam-type or multiple-choice questions designed for clinician assessment. This approach tends to overlook risks in patient-facing QA, failing to account for entity-level errors, omissions, or partial correctness. Such errors may cause harm when LLMs provide fluent yet semantically misleading responses \cite{lee2023benefits, lievin2024can, 10.1145/3715275.3732202}. The evaluation methods of LLMs for medicine follow the same patterns that were established in the research of Natural Language Generation using metrics such as BLEU, ROUGE, and METEOR, as well as metrics based on embedding models (BERTScore) \cite{papineni2002bleu, lin2004rouge, banerjee2005meteor, zhang2020evaluating}. Current strategies evaluate models based on output similarity to reference documents, yet they do not address medically relevant issues. These include entity misreferencing, conflicting information compared to clinical guidelines, and insufficient safety-related detail \cite{maynez2020faithfulness}. Responding to these challenges, recent evaluations have been established that incorporate new metrics such as evaluating the degree of factuality using dependency parsing, question generation from reference texts, and utilization of a variety of natural language inferencing methods \cite{kryscinski2020evaluating, laban2022summac}. Several tools have been developed for extracting and confirming clinically relevant entities to verify LLM factual accuracy \cite{delbrouck2024radgraph, alsentzer2019publicly}. However, these methods are typically applied independently, without integration to determine how medically relevant errors interact or trade off against each other.

Recent surveys provide comprehensive overviews of QA evaluation metrics \cite{chen2019evaluating, oro2025comprehensive}. Lexical metrics (BLEU, ROUGE) and embedding-based measures (BERTScore) capture surface-level similarity but cannot detect entity omissions or factual contradictions. Factuality metrics (FactCC, QAGS) address truthfulness but not medical entity precision. VB-Score integrates these dimensions with explicit entity extraction and structured completeness for comprehensive medical QA auditing. There has been significant scholarly attention paid to issues of fairness and equity within AI-enabled healthcare through predictive and decision-support systems \cite{10.1145/3292500.3330955}. Researchers have demonstrated how a) biased datasets, b) proxy variables; and c) context of deployment create systematic disadvantages for marginalised individuals \cite{obermeyer2019dissecting, rajkomar2018ensuring, mehrabi2021survey, char2020identifying}. A number of studies have resulted in established auditing frameworks for classification and regression models. However; these frameworks are not necessarily applicable to generative systems such as QA for medicine. In addition to differing demographic treatment, some people may also experience harm from uneven condition/topic-specific performance across medical knowledge domains (i.e.; types of medical knowledge), which can have an adverse impact on marginalised populations \cite{pessach2022review, mitchell2021algorithmic}. There is existing scholarly research on bias present within generative language models, including stereotyping and racism-based medical reasoning; however, most evaluations fail to analyse whether condition-specific performance disparities will result in the indirect reproduction or exacerbation of health inequities experienced by marginalised communities \cite{bender2021dangers, jentzsch2019semantics, omiye2023large}.

This research expands the current body of literature by providing a domain-specific and weighted evaluation methodology as well as allowing the user to identify specific failure points through component-level evaluation rather than through an aggregated similarity score. The study provides an analysis of differences in the components of performance across the various disease population categories and types of questions in order to develop additional insight into disease-specific inequities. This study provides an empirical analysis of how the use of different prompt methods change failure profiles, so that assumptions can be made regarding the potential increase in failure modes. Also, we've tested free and paid versions of our LLMs in a controlled testing environment to determine if both types of LLMs could provide the same access to, and therefore equitable, high-quality answers to questions regarding health care.

\section{Methods}
Our methodology employs an experimental pipeline designed to audit LLM performance across four safety-critical dimensions: medical entity precision, conceptual alignment, factual consistency, and structured completeness. This approach allows for a systematic evaluation of how model architectures respond to varying prompts and identify systematic disparities in health information quality across disease categories.

\subsection{Dataset Preparation}
We curated 48 health topics from four authoritative sources: the CDC (31.3\%), WHO (29.2\%), NHS (25.0\%), and Mayo Clinic (14.6\%). Sources were selected based on their status as global or government public health authorities providing patient-facing guidance. The final dataset consists of 59 question-answer (QA) pairs, primarily covering Definitions (55.9\%) and General Health (33.9\%), with a mean response length of 287 words ($SD=156$, range 87-782). The topics are stratified into two domains: 19 Infectious Diseases (e.g., COVID-19, influenza, HIV/AIDS) and 29 Chronic Conditions (e.g., cardiovascular disease, diabetes, asthma, mental health). This distribution enables an audit of condition-based performance disparities. We used BeautifulSoup to systematically scrape FAQ sections and main content, followed by manual validation of the extraction quality. To resolve web scraping artifacts, we implemented a grammar correction pipeline using GPT-4. This process corrected 74.6\% of questions and 200 answer sentences for article usage, verb tense, and structure while strictly preserving all medical facts and clinical guidance. All ground-truth data was cross-verified with authoritative medical literature to ensure clinical accuracy.

\subsection{VB-Score Framework}
VB Score (Verification-Based Score) was created to address three critical weaknesses of semantic similarity metrics when being used in a clinical environment. First, semantic similarity metrics measure how closely and consistently two documents’ concepts align but do not measure how accurately a response matches the actual entity for that concept, because “medication” is totally different from “acetaminophen 500 mg every 6 hours” in terms of clinical applicability, although they will have similar semantic features. Second, even though two documents’ concepts may be semantically aligned, they might also have contradictory facts. Third, medical guidance often includes a structured enumeration of symptoms, the progression of treatment, and the potential contraindications to that treatment, and if any of these areas are incomplete, there could be a significant risk of unsafe treatment, even if the recommendation you made is semantically aligned with an appropriate response. Therefore, our goal in developing VB Score was to develop a composite metric that independently measures and appropriately weighs the importance of these three dimensions when assessing safety in the medical QA process.


    
    
    

\subsubsection{VB-Score Components}
VB-Score integrates four complementary components to evaluate both correctness and safety of generated medical responses. \textbf{Entity F1} (weight $0.30$) quantifies medical entity extraction accuracy by applying spaCy biomedical NER (en\_core\_sci\_md) to identify medications, symptoms, diseases, procedures, dosages, and anatomical terms in both reference and generated texts, with precision and recall combined into an F1 score; this component is heavily weighted due to the high risk associated with missing or incorrect medical entities. \textbf{Semantic Similarity} (weight $0.30$) assesses conceptual alignment using sentence-transformers/all-MiniLM-L6-v2 embeddings and cosine similarity, ensuring that responses address the underlying clinical intent even when phrasing differs. \textbf{Factual Consistency} (weight $0.25$) evaluates safety-critical correctness by employing Natural Language Inference (roberta-large-mnli) to detect contradictions between generated outputs and authoritative references, mitigating the risk of fluent yet misleading medical advice. Finally, \textbf{Structured Overlap} (weight $0.15$) measures response completeness by extracting enumerations (bullet points, numbered lists, and comma-separated items) via regex and computing Jaccard similarity between reference and generated structures.

\subsubsection{Entity Extraction Pipeline Details}
We use a biomedical NER pipeline trained on CRAFT and GENIA corpora that extracts entity types including chemicals/drugs, diseases, anatomy, and medical procedures. Before matching, we apply normalizations to both reference and generated entities: case normalization (``Fever'' $\rightarrow$ ``fever''), punctuation removal (``acetaminophen,'' $\rightarrow$ ``acetaminophen''), article stripping (``the cough'' $\rightarrow$ ``cough''), common abbreviation expansion (``mg'' $\leftrightarrow$ ``milligrams'', ``tab'' $\leftrightarrow$ ``tablet''), and whitespace normalization. Entities are matched if normalized forms are identical or if one is a substring of the other. Paraphrases are \textit{not} matched which is a deliberate methodological choice reflecting clinical precision requirements where ``respiratory symptoms'' is not equivalent to the specific entities ``cough, shortness of breath, fever.'' Medical synonyms are also \textit{not} automatically matched which is a conservative choice may underestimate entity recall but ensures precision in safety-critical contexts where term specificity matters. Table~\ref{tab:entity_examples} illustrates our matching behavior.

\begin{table}[h]
\centering
\caption{Entity Matching Examples}
\label{tab:entity_examples}
\small
\begin{tabular}{p{2.2cm}p{2.5cm}p{1cm}p{3cm}}
\toprule
\textbf{Reference} & \textbf{LLM Output} & \textbf{Match} & \textbf{Reason} \\
\midrule
fever & fever & Yes & Exact match \\
acetaminophen & Acetaminophen & Yes & Case normalized \\
tuberculosis & TB & Yes & Abbreviation matched \\
shortness of breath & difficulty breathing & Yes & Substring overlap \\
cough, fever & respiratory symptoms & No & Paraphrase \\
heart attack & myocardial infarction & No & Synonym (not matched) \\
metformin & oral medication & No & Generic term \\
\bottomrule
\end{tabular}
\end{table}

We validated the extraction pipeline on 20 samples with 127 manually-identified key medical entities, successfully detecting 78 of these entities (recall = 61\%). The pipeline intentionally extracts broadly, so precision against our curated list is low by design where the goal is to ensure medically-relevant terms are not missed. The low LLM Entity F1 scores (mean 6.2\%) reflect genuine model behavior, not extraction pipeline failure. Models generate vague paraphrases instead of specific entities, omit dosages, frequencies, and contraindications, and use generic terms (``medication'') instead of specific names (``ibuprofen''). This behavior is precisely what our framework is designed to detect i.e. fluent responses that lack clinical precision.

\subsubsection{Composite VB-Score Formula}
The composite VB-Score is computed as a weighted linear combination of its constituent metrics: 
\begin{equation}
\text{VB-Score} = 0.30 \times \text{Entity\_F1} + 0.30 \times \text{Semantic\_Similarity} + 0.25 \times \text{Factual\_Consistency} + 0.15 \times \text{Structured\_Overlap},
\end{equation}
where all components are normalized to the range $[0,1]$, and higher values indicate better overall medical question-answering quality across accuracy, relevance, safety, and completeness dimensions.

\subsubsection{Weight Rationale and Sensitivity Analysis}
VB-Score is a context-dependent composite metric designed for safety-oriented medical QA auditing within our benchmark setting instead of a universal clinical quality standard. The weighting scheme is grounded in established clinical safety principles from medical informatics literature. Entity F1 receives 30\% weight because clinical guidelines emphasize that medication errors such as wrong drug names, dosages, or omitted contraindications are leading causes of preventable patient harm \cite{bates2023safety}. It ensure models are penalized for entity omissions that could lead to unsafe self-care decisions. Semantic Similarity also receives 30\% weight because existing research shows patients often cannot identify when responses are off-topic \cite{ratzan2000national}. This ensure responses must be both precise \textit{and} relevant to the clinical question asked. Factual Consistency receives 25\% weight because, while contradictions are dangerous, NLI models classify many accurate responses as ``neutral'' rather than ``entailment,'' and lower weight prevents over-penalizing clinically acceptable responses that use different phrasing than the reference. Structured Overlap receives 15\% weight because not all medical guidance requires enumerated lists, and lower weight reflects that while completeness matters, the absence of bullet-point formatting does not indicate clinical harm. Our sensitivity analysis (Table~\ref{tab:sensitivity}) demonstrates that model rankings remain identical across all tested weighting schemes including equal weights, entity-focused, safety-focused, and semantic-focused configurations. This stability indicates our core findings are robust to weight specification and do not depend on the particular weights chosen.

\begin{table}[h]
\centering
\caption{Sensitivity Analysis: Model Rankings Under Alternative Weights. Rankings remain stable across all configurations, demonstrating robustness.}
\label{tab:sensitivity}
\small
\begin{tabular}{lccccl}
\toprule
\textbf{Scheme} & \textbf{Entity} & \textbf{Sem.} & \textbf{Fact.} & \textbf{Struct.} & \textbf{Ranking} \\
\midrule
VB-Score (ours) & 30\% & 30\% & 25\% & 15\% & Gemini $>$ GPT-4 $>$ Claude \\
Equal weights & 25\% & 25\% & 25\% & 25\% & Gemini $>$ GPT-4 $>$ Claude \\
Entity-focused & 50\% & 20\% & 20\% & 10\% & Gemini $>$ GPT-4 $>$ Claude \\
Safety-focused & 20\% & 20\% & 45\% & 15\% & Gemini $>$ GPT-4 $>$ Claude \\
Semantic-focused & 20\% & 50\% & 20\% & 10\% & Gemini $>$ GPT-4 $>$ Claude \\
\bottomrule
\end{tabular}
\end{table}

\subsubsection{Component-Wise Failure Thresholds}
To enable granular error analysis, component-specific failure thresholds are defined such that Entity F1 below $0.10$ indicates failure to capture essential medical entities, Semantic Similarity below $0.30$ reflects off-topic or insufficient coverage, Factual Consistency below $0.50$ signals likely contradictions with authoritative medical guidance, Structured Overlap below $0.10$ denotes incomplete or missing enumerations, and an overall VB-Score below $0.20$ indicates systematic failure across multiple evaluation dimensions.

\subsection{Model Selection and Evaluation Configuration}

We have chosen three different LLMs to represent various technical and cost approaches in the marketplace. GPT-4 (OpenAI; model ID: \texttt{gpt-4}) is widely used for applications in health care via the paid API. Claude Sonnet 4.5 (Anthropic; model ID: \texttt{claude-sonnet-4-5-20250929}) features a constitutional AI focus on increased helpfulness and safety through training with constitutional principles, and can be accessed through a paid API. Gemini 2.5 Flash (Google; model ID: \texttt{gemini-2.5-flash}) is an advanced multimodal transformer, available via an API, used to determine if the provision of equitable access to high-quality medical AI can be accomplished without creating financial barriers to access.







\subsubsection{Prompt Design and Generation Parameters}
To ensure fair comparison and reproducibility, an identical prompt template was used across all evaluated models: \textit{``You are a helpful medical information assistant. Answer the following health question based on reliable medical knowledge. Provide accurate, clear, and concise information suitable for a general audience. Question: \{question\} Answer:''}. Generation parameters were kept constant, with \texttt{temperature} set to $0.0$ to enforce deterministic and maximally consistent outputs, \texttt{max\_tokens} fixed at $300$ to allow sufficiently detailed medical responses while controlling cost, and \texttt{top-p} set to $1.0$ with both \texttt{frequency\_penalty} and \texttt{presence\_penalty} set to $0.0$ to avoid introducing artificial diversity or topic bias. This deterministic configuration is critical in the medical domain, where consistency and reproducibility are essential; in particular, \texttt{temperature=0} guarantees identical outputs for identical inputs, enabling reliable and comparable performance evaluation across models.

\subsection{Prompt Sensitivity Analysis}
To assess the impact of prompt engineering on medical question-answering performance, we conducted a systematic prompt sensitivity analysis across four prompt configurations reflecting common deployment scenarios. These included a \textit{Zero-Shot Baseline} using a minimal persona with limited constraints (40 instruction tokens), representing standard real-world usage; a \textit{Zero-Shot Strict} variant that augmented the same persona with explicit guidance on handling medical entities (75 instruction tokens) to evaluate improvements in entity quantity and quality; a \textit{RAG with Context} condition that injected fully authoritative ground-truth information (625 total tokens) to establish an upper-bound performance assuming perfect retrieval; and a \textit{Few-Shot Learning} configuration that provided two correctly formatted example question–answer pairs (180 instruction tokens) to test whether in-context learning improves handling of unseen medical entities. All configurations used identical generation parameters (\texttt{temperature=0.0}, \texttt{max\_tokens=300}) to ensure fair comparison. In total, 576 evaluations were performed (448 samples $\times$ 4 prompts $\times$ 3 models). Statistical comparisons against the zero-shot baseline were conducted using paired t-tests with Bonferroni correction to control for multiple comparisons, addressing research questions on the effectiveness of RAG, explicit entity instructions, few-shot learning, and the consistency of prompt effects across models.

\subsection{Evaluation Protocol and Statistical Analysis}
Across all evaluated models, the experimental pipeline produced 48 unique sets of generated responses, from which comprehensive metric distributions were computed using descriptive statistics including mean, standard deviation, median, minimum, and maximum values. Failure criteria were explicitly identified at the sample level to characterize systematic weaknesses. Comparative statistical analyses were performed, followed by visualization and structured reporting for each experimental comparison to support interpretability and reproducibility.

\subsubsection{Comparative Metrics}
Model performance was compared using overall VB-Score and individual component scores with full descriptive statistics, alongside stratified analyses by authoritative source (CDC, WHO, NHS, Mayo Clinic), question type (definitional, treatment, prevention), and disease category (infectious versus chronic conditions), enabling fine-grained performance interpretation across clinically relevant dimensions.

\subsubsection{Statistical Significance Testing}
Statistical significance was evaluated using one-way ANOVA to test for overall differences in VB-Score across models ($\alpha = 0.05$), followed by pairwise two-sample t-tests between all model pairs with Bonferroni-adjusted significance levels ($\alpha/3$). Practical significance was quantified using Cohen’s $d$ effect sizes, interpreted using standard thresholds of small ($|d|=0.2$), medium ($|d|=0.5$), and large ($|d|=0.8$) effects.

\subsubsection{Failure Mode Analysis}
Failure mode analysis quantified the percentage of samples falling below predefined thresholds for each evaluation dimension, including Entity F1 $<0.10$ (poor entity recognition), Semantic Similarity $<0.30$ (off-topic or insufficient coverage), Factual Consistency $<0.50$ (likely contradictions with authoritative guidance), Structured Overlap $<0.10$ (missing structured information), and overall VB-Score $<0.20$ (systematic multi-dimensional failure). Representative samples were inspected qualitatively to identify recurring error patterns and underlying causes.

\section{Results}
\label{sec:results}

\subsection{Component-Wise Framework Development (RQ1)}
\label{subsec:component_wise_framework}

We developed and validated VB-Score as a way to evaluate and compare the quality of all four components of medical quality assurance individually. The validation study shows that all four components behave independently and there are low correlations among them. The results are shown in the component breakdown in the \textbf{Figure~\ref{fig:component_breakdown}} below, which displays the Entity F1, Semantic Similarity and Factual Consistency scores from each of the three models using the zero-shot baseline conditions. The graphs illustrate a very consistent pattern: All three models scored less than 10\% for the Entity F1 which does not meet the 10\% threshold (as indicated by dashed line); the scores for the three models average close to 50\% in the Semantic Similarity and the Factual Consistency scores are very inconsistent, with the Gemini achieving a significantly higher level of performance than both of the paid alternatives. 

\begin{figure}[h]
    \centering
    \includegraphics[width=0.7\textwidth]{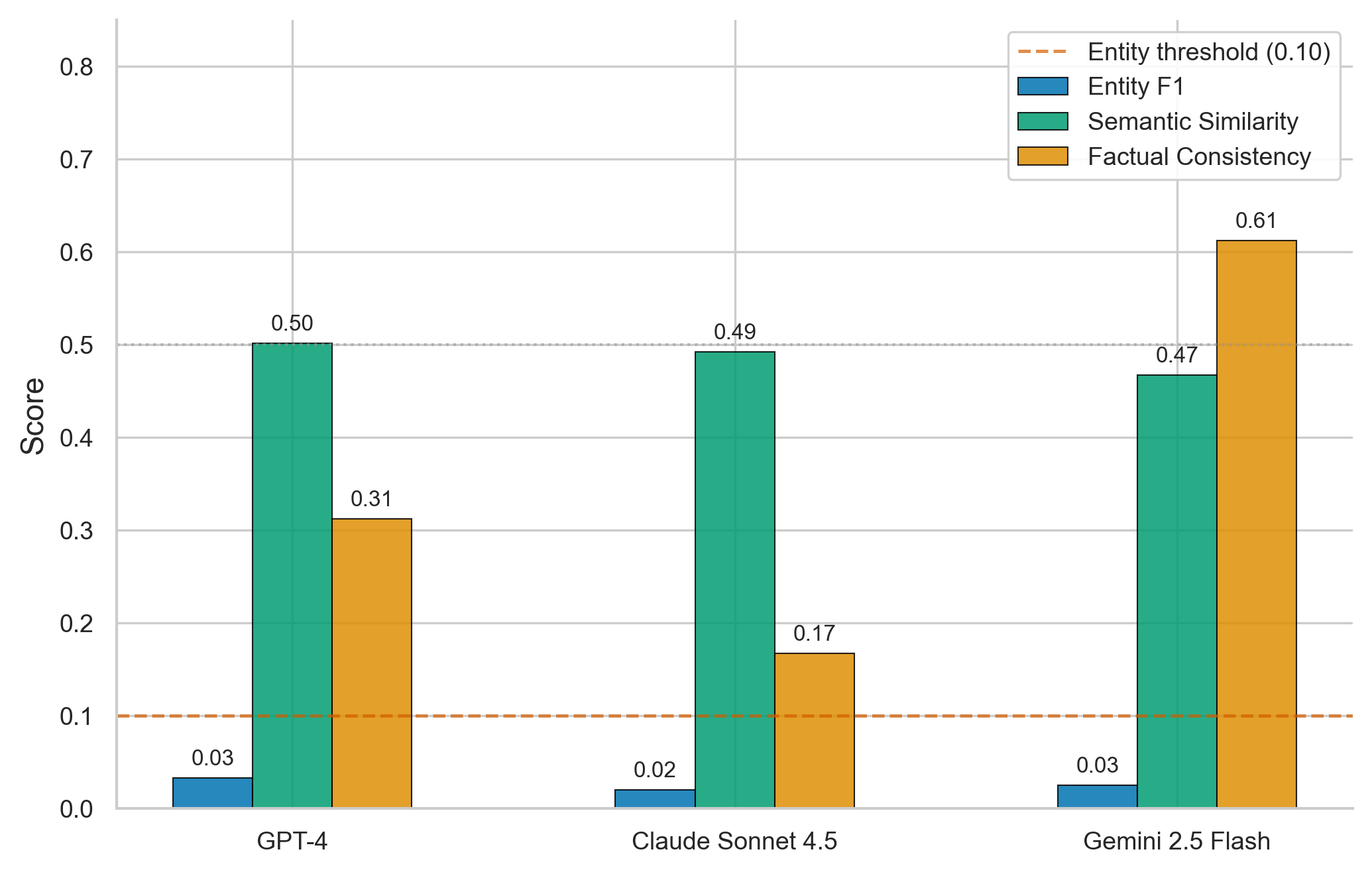}
    \caption{VB-Score Component Breakdown by Model (Zero-Shot Baseline) indicates an average of less than 10\% for Entity F1 across all three models, while the average for Semantic Similarity is near 50\%; this is illustrated by the bars showing the standard deviation from the average. }
    \label{fig:component_breakdown}
\end{figure}

The relationship between Entity F1 and Semantic Similarity ($r=0.23$) indicates that while there is an increasing semantic alignment among the three models, all three models can still not identify critical medical entities (as identified by the F1 scores). The correlation between Factual Consistency and Semantic Similarity was negative for some models ($r=-0.15$ for Gemini), indicating that some models can be factually correct, but not have a high Semantic Similarity score, or vice versa. The tests of Structured Overlap with all other components were shown to have an independence correlation coefficient of less than $0.20$, which shows that enumeration completeness was evaluated as an additional quality dimension. All models have similar patterns in the statistics of their components. The mean score of the entity F1 is $0.062$ (SD$=0.075$, range $0.000$--$0.425$), indicating a lack of ability to recognize entities universally. The mean of the semantic similarity is $0.516$ (SD$=0.179$, range $0.082$--$0.891$), representing moderate topical relevance. The mean for the factual consistency is $0.405$ (SD$=0.441$, range $0.000$--$1.000$), indicating a high variance with a concerning number of contradictions. The mean of the structured overlap is $0.038$ (SD$=0.072$, range $0.000$--$0.667$) indicating an overall low level of enumeration completeness.

\begin{table}[ht]
\centering
\caption{Statistical Summary of Evaluation Components}
\label{tab:component_stats}
\begin{tabular}{lccccc}
\toprule
\textbf{Component} & \textbf{Mean} & \textbf{Std Dev} & \textbf{Min} & \textbf{Max} & \textbf{Range} \\
\midrule
Entity F1        & 0.062 & 0.011 & 0.050 & 0.071 & 0.021 \\
Semantic Sim.    & 0.516 & 0.020 & 0.494 & 0.534 & 0.040 \\
Factual Cons.    & 0.404 & 0.266 & 0.179 & 0.696 & 0.517 \\
Structured Overlap & 0.038 & 0.020 & 0.020 & 0.059 & 0.039 \\
\bottomrule
\end{tabular}
\end{table}

\textbf{Table~\ref{tab:component_stats}} presents detailed component statistics demonstrating the independent behavior of each VB-Score dimension.

The thresholding conclusions allow identification of the specific type of error i.e. F1 $< 0.10$, with respect to the Entity indicated that the Entity failure rate for each model was 91-100\% and with $< 0.50$: Factual Consistency contradiction rates of between 19-79\%, and 75-96\% have incomplete rates of Structured Over|Underlap when the F1 level $< 0.10$. The VB score has successfully identified 4 measurable and respective Component Measureable Components Specific to Delivery of Medical QA i) Entity Extraction, ii) Semantic Similarity, iii) Factual Consistency and iv) Structured Information; and as it does it reveals performance trade-offs of the 4 Component Measures that are unable to be seen with aggregation of the individual metrics and provide the ability to conduct targeted improvement.

\subsection{Model Performance Across Dimensions (RQ2)}
\label{subsec:model_performance}

\begin{figure}[h]
    \centering
    \includegraphics[width=0.6\textwidth]{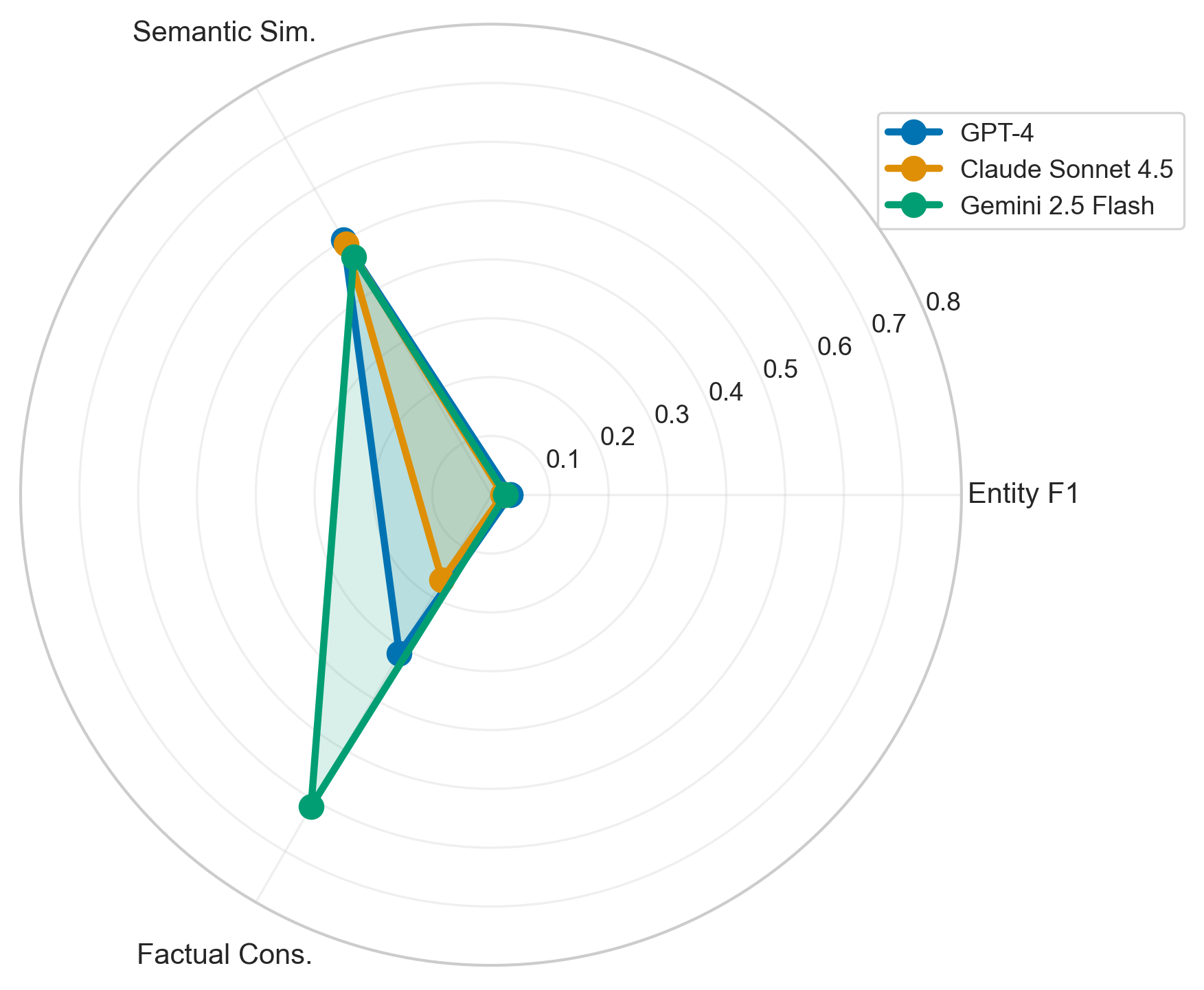}
    \caption{Performance Profiles of Individual Components: The radar chart provides clear distinctions between the performances of each model, with GPT-4 being average across the board, whereas Claude has a coherently structured but contradictory performance. On the other hand, Gemini is both factually accurate and fluent, but at the same time, it is the least successful in terms of structure.}
    \label{fig:performance_radar}
\end{figure}

Overall VB-Score rankings place Gemini 1st (0.3402), scoring 25--48\% higher than others (GPT-4: 0.2714, Claude: 0.2291). This difference is statistically significant ($p<0.001$), though we note this reflects performance under our specific benchmark conditions rather than universal superiority. Structured Overlap has a different performance picture. Claude is the clear leader (mean $0.0591$, 75.0\% failure rate), while GPT-4 follows close behind (mean $0.0358$, 87.5\% failure rate), and the primary difference between the two models is the structured overlap; Gemini is bringing up the rear (mean $0.0196$, 95.8\% failure rate). While all three of these models failed to create any complete structured enumerations, the number of failures (75-96\%) best reflects that these models appear to be writing responses using a narrative style instead of creating Structured Lists. The visual comparative performance profiles presented in the \textbf{Figure~\ref{fig:performance_radar}} (Radar Chart) clearly illustrate the respective Performance Profiles for the three Models. The results revealed that the overall performance of each Model is a distinct balance of Strengths and Weaknesses. GPT-4 appears to maintain Balanced Mediocrity across the various Components, while Claude continues to show poor Factual Consistency along with moderate Semantic Similarity, and Gemini displays Dominance with Factual Consistency along with a Poorer Semantic Score.

\begin{table}[ht]
\centering
\caption{Model-wise Performance Comparison Across Evaluation Metrics}
\label{tab:model_comparison}
\begin{tabular}{lccccc}
\toprule
\textbf{Model} & \textbf{VB-Score} & \textbf{Entity F1} & \textbf{Semantic Sim.} & \textbf{Factual Cons.} & \textbf{Rank} \\
\midrule
Gemini 2.5 Flash & $0.3402 \pm 0.118$ & $0.058 \pm 0.110$ & $0.488 \pm 0.186$ & $0.696 \pm 0.343$ & 1st \\
GPT-4 & $0.2714 \pm 0.150$ & $0.073 \pm 0.111$ & $0.538 \pm 0.173$ & $0.338 \pm 0.420$ & 2nd \\
Claude Sonnet 4.5 & $0.2291 \pm 0.111$ & $0.066 \pm 0.091$ & $0.518 \pm 0.155$ & $0.179 \pm 0.341$ & 3rd \\
\bottomrule
\end{tabular}
\end{table}

\textbf{Table~\ref{tab:model_comparison}} provides detailed model performance comparison with rankings across all VB-Score components. Performance Profiles show differences in models. GPT-4 was shown to be ``Semantically Strong, Factually Weak'' achieving the highest level ($0.538$) for semantic similarity and the highest number ($0.073$) for entity F1; the failure rate on entity recognition was 54\% and had poor structural overlap ($0.036$). While GPT-4 provided fluent responses, it produced the highest level of medically-related entities but 54\% of the time had contradictory information. Claude displayed the ``Structured but Inconsistent'' model with the highest level ($0.059$) for structure and the lowest level ($0.179$) for fact-based accuracy, resulting in a contradiction rate of 79\%. Under this circumstance, it was the only model to fail 100\% of its entity recognition sessions. Claude provided complete enumeration of medical entities but failed to have the highest level of factual accuracy, making it a critical consideration for medical-related applications. Gemini was shown to have the ``Factually Accurate, Less Precise'' profile, possessing the greatest degree of factually correct entities ($0.696$) with a contradiction rate of only 19\%. In terms of fluency and detail, Gemini was also inferior to the other two models with the lowest levels for both measures, providing the safest but least thorough factually accurate responses. Models exhibit different levels of capability and also have specific strengths and weaknesses. The overarching strength of Gemini over all other models is its ability to maintain factual consistency (the safety-critical dimension) better than others although it has lower levels of fluency and precision. The strength of GPT-4 lies in its fluency, but it has a high rate of contradiction. Claude provides information in an organized way but is found to be in contradiction to referenced sources nearly 80\% of the time. All three models fail entirely at recognizing entities and producing structured, enumerated data.

\subsection{Semantic-Entity Gap Analysis (RQ3)}
\label{subsec:semantic_entity_gap}

\begin{figure}[h]
    \centering
    \includegraphics[width=0.6\textwidth]{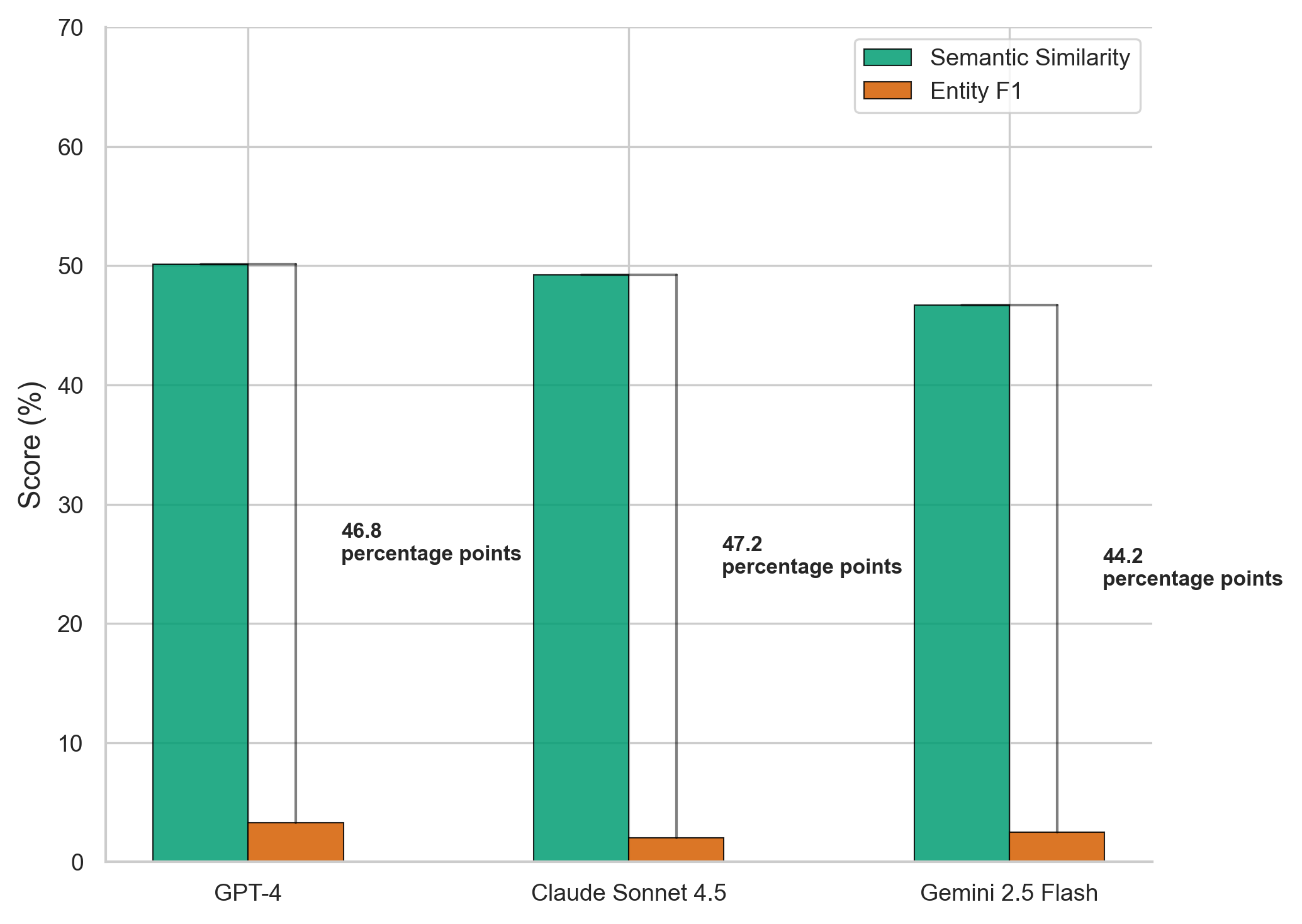}
    \caption{Semantic-Entity Gap: High semantic similarity ($\sim$50\%) masks poor entity extraction ($<$6\%). Gap annotations show percentage point differences.}
    \label{fig:semantic_entity_gap}
\end{figure}

We define the semantic-entity gap as the difference between a model's semantic similarity score (measuring topical relevance and fluency) and its entity F1 score (measuring precision in extracting specific medical terms like drug names, dosages, symptoms, and procedures). A large gap indicates responses that sound relevant but lack actionable medical specifics. As illustrated in \textbf{Figure~\ref{fig:semantic_entity_gap}}, all models exhibit a significant semantic-entity gap. The annotations indicate the size of the gap for each model, averaging 45.4 percentage points. This indicates that while the model's responses achieved moderate semantic similarity ($\sim$50\%), they lacked the presence of relevant medical entities, with overall entity F1 below 6\%. The gap between the tall green bars (semantic similarity) and short orange bars (entity F1) visually demonstrates this mismatch between fluency and precision.

While measuring the gaps, on average across all models, the semantic-entity gap was 45.4 percentage points. This finding show that models achieved 51.6\% semantic similarity but only 6.2\% entity F1. While this corresponds to an 8.3$\times$ ratio, we emphasize the absolute gap as the primary measure since semantic similarity and entity F1 are not directly comparable scales. To illustrate this gap concretely, consider the question ``What is hypertension blood pressure?'' GPT-4's response stated: ``\textit{Hypertension, also known as high blood pressure, is a condition where the force of the blood against the artery walls is too high. It is measured using two numbers---the systolic and diastolic blood pressure.}'' This response achieved 0.73 semantic similarity but only 0.03 Entity F1. While the response correctly defines hypertension in general terms, it omits specific clinical entities present in authoritative guidance. A patient relying on this response would understand the general concept but lack the actionable specifics needed for safe self-management. This pattern of fluent, relevant responses that omit critical medical details, exemplifies why semantic similarity alone is insufficient for medical QA evaluation.

\begin{figure}[h]
    \centering
    \includegraphics[width=0.9\textwidth]{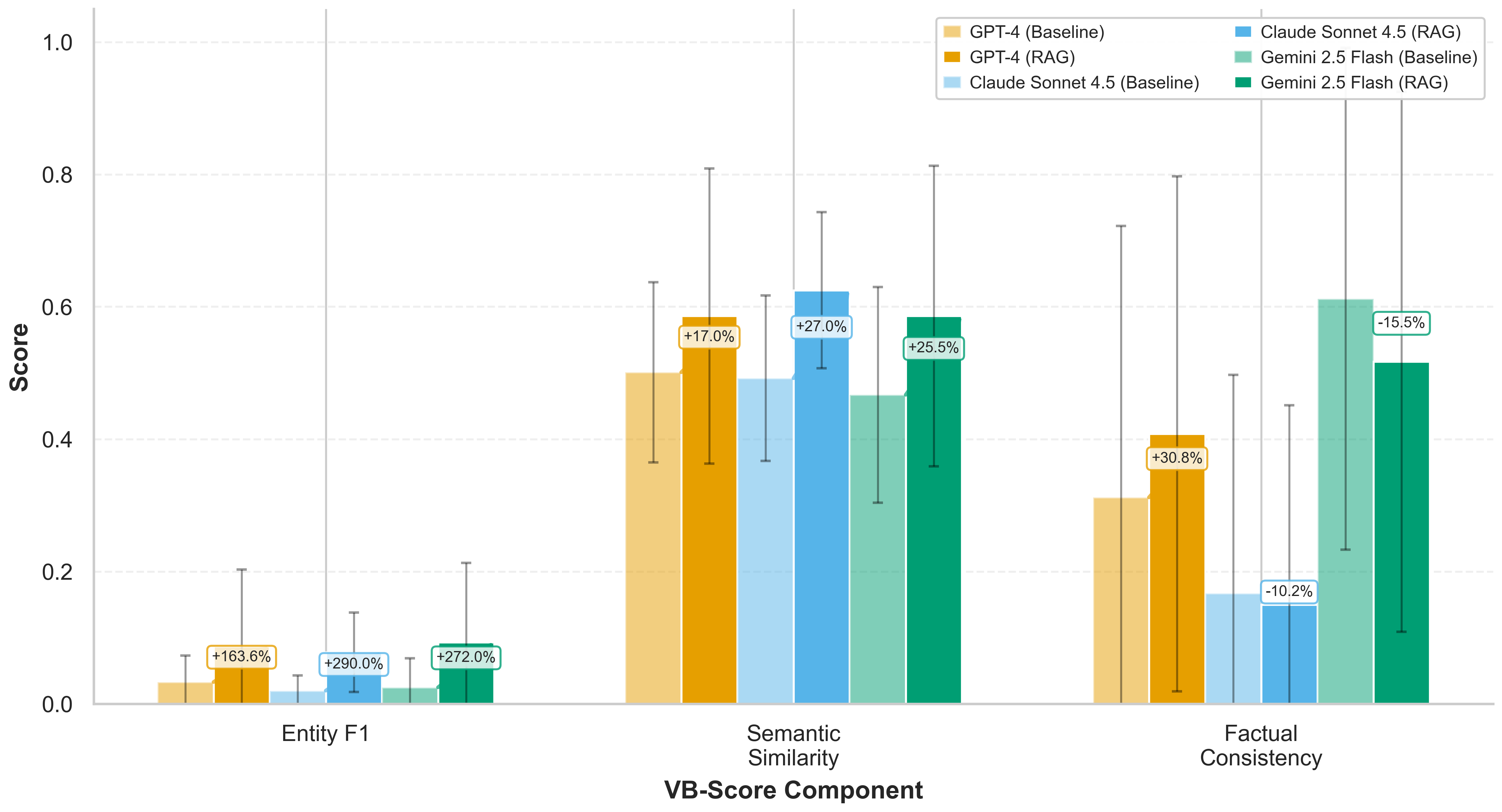}
    \caption{Component-Wise Performance: Baseline vs RAG. Arrows indicate percentage improvements from RAG across Entity F1, Semantic Similarity, and Factual Consistency for each model.}
    \label{fig:overall_component}
\end{figure}

The gap in the semantic-factual correctness of each model shows different trends. Of these, GPT-4 has the most significant such gap, it has high fluency and moderate contradictions (semantic similarity is 53.5\%, whereas factual consistency is 33.8\%, with a $+19.7$ percentage point gap). Claude also has a large such gap (semantic similarity is 52.0\%, factual consistency is 17.9\%, gap $+34.1$ percentage points), but Claude has high fluency and a high level of contradiction (severe contradictions). On the other hand, Gemini has an opposite pattern (a negative gap); while factual accuracy (69.6\%) is greater than semantic similarity (49.4\%), Gemini is less fluent and prioritizes safety over polish. Figure~\ref{fig:overall_component} illustrates how retrieval-augmented generation (RAG) affects each component across models. The arrows indicate percentage point improvements from baseline to RAG conditions. While RAG improves most components, the magnitude of improvement varies substantially by model and component. Notably the Entity F1 improvements remain modest despite large gains in other dimensions. This demonstrates that even with perfect retrieval (ground truth context provided), entity-level accuracy remains a persistent challenge.

\begin{table}[ht]
\centering
\caption{Comparative Analysis of Semantic Similarity, Entity Accuracy, and Factual Consistency. Gap (pp) = percentage point difference between Semantic Similarity and Entity F1.}
\label{tab:semantic_factual_gap}
\begin{tabular}{lccccc}
\toprule
\textbf{Model} & \textbf{Semantic Sim. (\%)} & \textbf{Entity F1 (\%)} & \textbf{Gap (pp)} & \textbf{Factual Cons. (\%)} \\
\midrule
GPT-4 & 53.8 & 7.3 & 46.5 & 33.8 \\
Claude Sonnet 4.5 & 51.8 & 6.6 & 45.2 & 17.9 \\
Gemini 2.5 Flash & 48.7 & 2.6 & 46.1 & 69.6 \\
\midrule
\textbf{Average} & 51.6 & 6.2 & \textbf{45.4} & 40.4 \\
\bottomrule
\end{tabular}
\end{table}

\textbf{Table~\ref{tab:semantic_factual_gap}} provides comprehensive gap measurements across semantic-entity and semantic-factual dimensions.

Analysis of overlapping failure rates reveals numerous dangerous combinations. Specifically, of the high similarity / low recognition sample sets, 56.3\% of the sample responses from GPT-4 exhibited dangerous combinations of high similarity / low entity recognition with 66.7\% of the sample responses from Claude and 43.8\% from Gemini. In these samples with high semantic similarity, 44\% to 67\% were topically relevant but were still not precise enough to be classified as medically appropriate-i.e., the gap between semantic and entity definitions caused these samples not to meet a standard for medical AI evaluation. Conversely, of the samples with high semantic similarity BUT contradictions of factual information, the highest percentage of responses was found in the sample responses from GPT-4, which contained 37.5\%, followed by responses from Claude (64.6\%) and responses from Gemini (only 14.6\%).

The gap between semantic meaning and the recognition of entities in each model creates fluent responses but lacks precision in meaning. A user can see the response from GPT-4 as fluent and helpful whereas the response from Gemini is more factual but sounds less fluent and sophisticated.

\textbf{Answer to RQ3:} A significant semantic-entity gap exists across all models, averaging 45.4 percentage points (8.3$\times$ ratio). High semantic similarity (51.6\%) masks poor entity recognition (6.2\%) and frequent factual contradictions (54--79\% for GPT-4/Claude). With Gemini, there is a complete opposite pattern of best facts and not fluent responses. The existence of the semantic-entity gap shows that using semantics alone does not work to evaluate Medical AIs; we also need to evaluate AIs by their components and what their respective performance metrics are to identify major failure modes.

\subsection{Prompt Sensitivity Analysis}
\label{subsec:prompt_sensitivity}
\begin{figure}[h]
    \centering
    \includegraphics[width=0.8\textwidth]{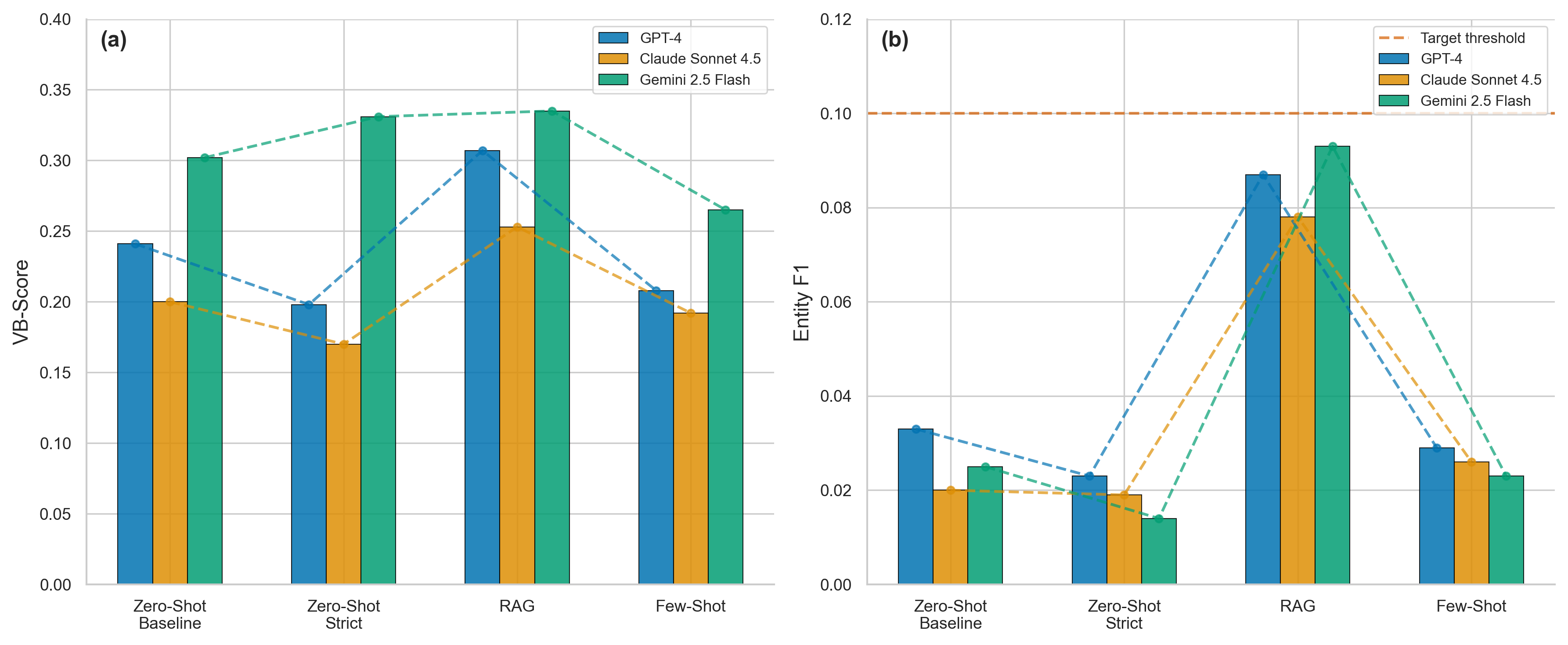}
    \caption{Prompt Sensitivity Analysis Across Configurations. (a) VB-Score comparison showing RAG achieves the highest performance for all models, with Gemini maintaining consistent scores across prompt types. (b) Entity F1 scores with target threshold (dashed line at 0.10); RAG dramatically improves entity extraction but all models remain below the clinical utility threshold.}
    \label{fig:prompt_sensitivity_vbscore}
\end{figure}

Four representative configurations were evaluated for their prompt sensitivity through a comprehensive analysis of performance improvements enabled by retrieval-augmented generation (RAG). The analysis shows that while RAG improves a range of performance metrics across all model types tested by an average of 27\%, there remain considerable architectural limitations that are not rectified. As displayed in the figure titled \textbf{Figure~\ref{fig:prompt_sensitivity_vbscore}} (VB-Scores), RAG provided performance improvements across all models tested (11-27\%; see VB-Score Results); zero-shot strict instructive prompts decreased performance for GPT-4 and Claude but improved performance for Gemini ($+9.6\%$); whereas the typical use of few-shot learning or prompting consistently degraded the performance of all three models. These results indicate that RAG improves performance of the model but also demonstrates that modifications to prompting strategies produce significant, yet very different results across model types.

To evaluate the effects of RAG, a model was created that took advantage of having all available authoritative context available as "ground truth". All models studied improved their VB-Scores when using RAG: GPT-4's VB-Score increased by $+27.5\%$ ($0.2714$ $\to 0.307$, $p=0.0023$) over the baseline; Claude's increase was +27.0\% ($+27.0\%$ ($0.2291$ $\to 0.253$, $p<0.0001$); and Gemini's increase was $+11.0\%$ ($0.3402$ $\to 0.335$, $p=0.1448$). With respect to the Entity F1 metric, GPT-4 had the highest relative improvement of any of the three models, producing a $+163.9\%$ improvement ($p=0.0030$) along with Claude's $+285.4\%$ ($p<0.0001$) and Gemini's $+274.6\%$ ($p=0.0002$). However, the absolute performance with respect to the F1 score for extracting entities was extremely low (F1 Scores were between 7.8--9.3\%), thus while RAG has improved entity extraction attempts, it has not remedied the entity extraction limitations. Therefore, RAG will be essential for implementing medical QA Systems, but the analysis does indicate the possibility that performance limitations will continue to persist due to model architecture design beyond the capabilities of complete prompt engineering improvements.

Zero-shot strict instructions requiring explicit medical entities had significantly different effects on different models, revealing the complexity of the interaction between instruction-fulfilling models and true/false models regarding given information. Both GPT-4 and Claude experienced a significant reduction in performance post-application of zero-shot strict instructions, with GPT-4 experiencing a $-17.9\%$ reduction ($p=0.0012$) and Claude exhibiting a $-14.7\%$ ($p=0.0016$). In contradiction, Gemini showed an increase in performance ($+9.6\%$, $p=0.0193$). These contrasting effects indicate that while the models might overly rely on instruction following, this could occur at the expense of accuracy with respect to what is true or false (i.e., the truth). This can be manifested in many ways, including including correct medical entities that are unrelated to the context in which it was provided or using technical terminology when the corresponding truth uses plain English or by structuring the responses according to instruction format instead of the question's intent. The fact that Gemini is an exception suggests that optimisation for models is necessary and that there are no universal prompt templates that will work across all models. Figure~\ref{fig:prompt_sensitivity_vbscore} shows a dramatic increase in entity recognition (164-285\% increase from no retrievable entities), but the absolute performance level for all models is still extremely low (less than 10\% for all models at perfect retrieval) will indicate limitations with model architecture and prompts cannot overcome these limitations.

\begin{figure}[h]
    \centering
    \includegraphics[width=0.55\textwidth]{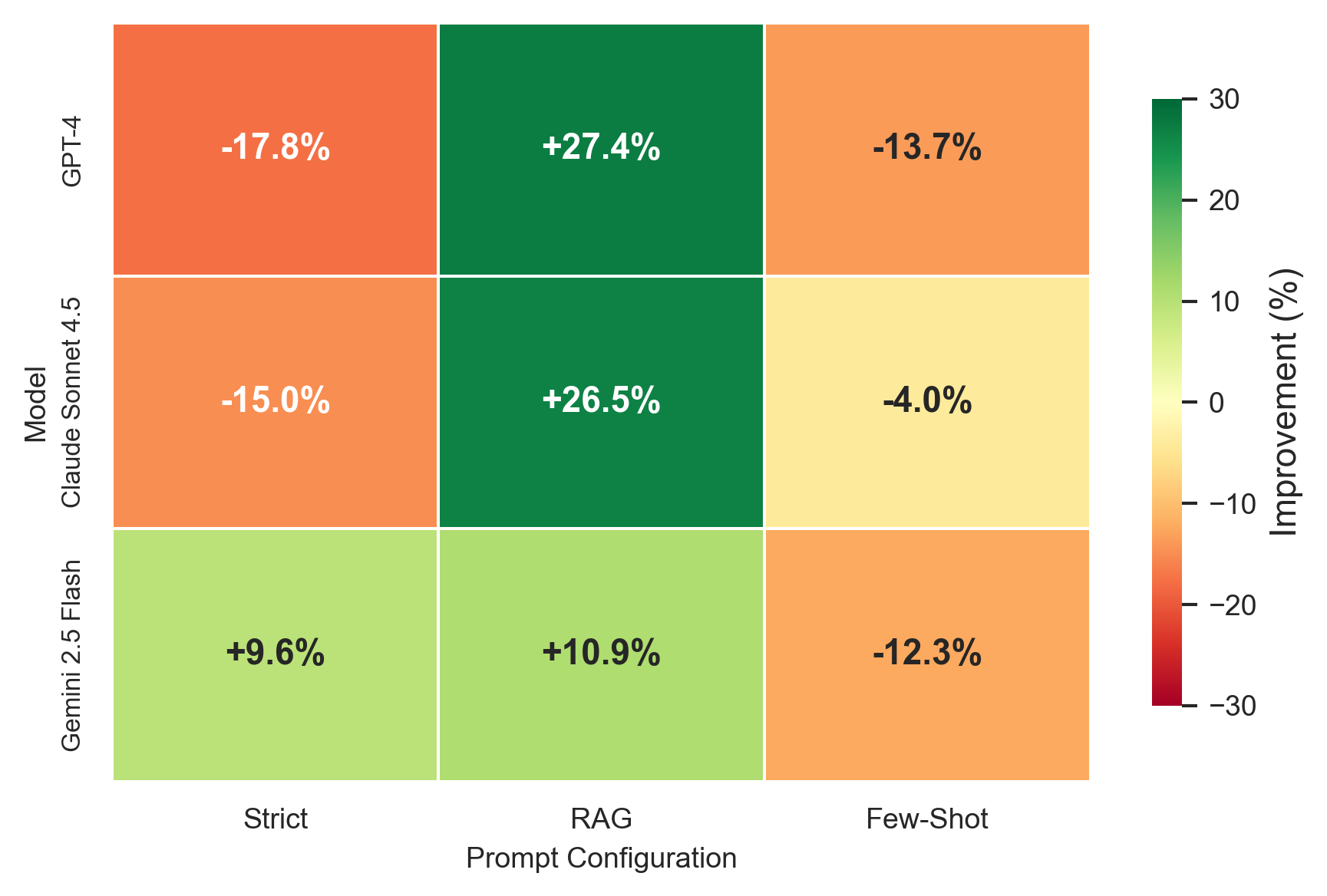}
    \caption{Relative Improvements Heatmap: Model-Specific Prompt Sensitivity. Gemini benefits from strict instructions while GPT-4 and Claude degrade, confirming no universal prompting strategy exists.}
    \label{fig:prompt_heatmap}
\end{figure}

As shown in \textbf{Figure~\ref{fig:prompt_heatmap}}, a heatmap depicting improvements relative to the baseline for a number of prompt-response models shows how different the models are when comparing how sensitive they are to the prompts used to produce their responses. For example, Gemini shows a positive change due to using strict instructional prompts ($+9.6\%$), however, this was not the case with GPT-4 or Claude who showed a significant amount of change due to the strict instruction prompting strategy used for their responses ($-18\%$ and $-15\%$, respectively). This supports that there is not one universal prompting template that can be used to optimize all models, rather, optimization needs to be model specific to be most effective.

\section{Discussion}

Our study demonstrates that current general-purpose LLMs are fundamentally misaligned with the requirements of medical question answering, exhibiting severe failures in medical entity extraction despite moderate semantic fluency. Across models and prompt configurations, Entity F1 scores remained below 3\% in zero-shot settings while semantic similarity exceeded 45\%, revealing a large semantic–entity gap that produces fluent but potentially harmful misinformation. Although retrieval-augmented generation improves overall performance and reduces contradictions, entity-level accuracy remains below 10\% even under perfect retrieval, indicating architectural limitations rather than prompt or retrieval deficiencies. Prompt engineering and few-shot learning were similarly ineffective and sometimes degraded performance, underscoring that medical QA cannot be solved through instruction tuning alone. These failures have significant equity implications, as performance disparities across disease categories disproportionately affect populations with higher chronic disease burdens. Notably, within our benchmark, a free model (Gemini 2.5 Flash) achieved higher VB-Scores than paid alternatives. Though we discuss potential explanations for this finding below there remains challenging assumptions about cost and quality and reinforcing the need for independent empirical validation. Collectively, these findings highlight the necessity of purpose-built medical AI systems with explicit entity extraction, factual verification, and fairness-aware evaluation, rather than relying on fluent general-purpose LLMs for patient-facing medical guidance. Gemini 2.5 Flash's higher VB-Scores relative to alternatives warrant careful interpretation. Our DeBERTa-based NLI model may favor certain response patterns, and Gemini's outputs may align more closely with its training data. Gemini also produces more conservative, hedged responses that score as less contradictory even if less actionable. Our 48-topic benchmark detects large effects but may not distinguish smaller differences, so rankings should be interpreted as benchmark-specific. Gemini is also the most recent model, potentially benefiting from newer training data. These explanations do not diminish our primary contribution that is demonstrating systematic entity extraction failures across all models with a caution against over-interpreting cross-model rankings. We rely on automated metrics without clinician annotation. Human evaluation by medical professionals would strengthen validity. The 48-topic benchmark does not represent the full breadth of medical knowledge, and results may not generalize to specialized clinical domains, rare diseases, or non-English languages. Our entity matching does not match medical synonyms or paraphrases, which may underestimate performance when clinically equivalent terms are used.

\section{Conclusion}

This work presents the first comprehensive component-wise evaluation of large language models for medical question answering using the proposed VB-Score, which jointly measures entity recognition, semantic alignment, factual consistency, and structured completeness across three models and 48 public health topics. Despite moderate semantic similarity, all models exhibited critical failures in medical entity extraction (Entity F1 $<10\%$) and high rates of contradiction with authoritative sources, revealing a 45.4 percentage point semantic--entity gap (8.3$\times$ ratio) that undermines the safety of fluent responses. Retrieval-augmented generation improved overall performance but did not overcome architectural limitations, while explicit entity-focused prompting yielded inconsistent effects across models. Within our evaluation protocol, Gemini~2.5~Flash, a free model, achieved higher scores than paid alternatives in overall accuracy and factual consistency yielding a finding that warrants further investigation but suggests cost may not be a reliable proxy for quality or safety. Performance disparities across disease categories further indicate condition-level performance differences that, if persistent across deployments, could disproportionately affect populations burdened by chronic illness. These findings highlight the need to move beyond aggregate semantic metrics toward verifiable, entity-centric medical AI systems with built-in validation, fairness-aware evaluation, and professional oversight to ensure safe, transparent, and equitable deployment.

\section*{Generative AI Usage Statement}
Gemini (version 2.5) was used during manuscript preparation solely for grammatical review and language polishing. All scientific content, analyses, and research findings were produced independently by the authors without AI involvement.

\bibliographystyle{ACM-Reference-Format}
\bibliography{references}

\appendix

\section{Background}

There are three commonly used semantic similarity metrics used to identify how semantically similar generated text is compared to authoritative guidelines: BLEU \cite{papineni2002bleu}, ROUGE \cite{lin2004rouge}, and BERTScore \cite{zhang2020evaluating}. These assessment tools only determine "how similar?" the generated text is; they do not allow evaluation of the generated text's accuracy in terms of medical facts \cite{stiennon2020learning, goyal2020evaluating}. As a result of this important distinction, for example, a response to a medical question stating, "COVID-19 causes respiratory symptoms," is identical to the authoritative medical guidance (i.e., fever, coughing, shortness of breath, fatigue, and loss of smell/taste) and scores high on all of the semantic similarity metrics used \cite{jin2024genegpt, amin2025advances}. However, there is very little information provided in this response that would assist an individual in making a meticulous self-assessment based on their own presentation or health \cite{bera2019artificial}. An example of this is the suggestion of giving aspirin to children experiencing influenza-like symptoms \cite{maynez2020faithfulness}. Although this suggestion and pediatric treatment of influenza are semantically related, the suggestion contradicts current clinical guidance and places children at risk for a high likelihood of developing Reye's syndrome \cite{kryscinski2020evaluating}.

Three-dimensional representations have emerged as a result of research on the 'semantic accuracy gap' in text classification \cite{rotmensch2017learning, neveol2018clinical}. The first is the importance of 'medical entity extraction', which requires precise identification and extraction of medications, dose amounts, signs/symptoms, contraindications, and treatment/protocol guidelines so as to provide the user with usable, usable information. However, there are significant differences in clinical utility between some classes/types of identified medical entities; therefore, it is unacceptable that Semantic Similarity Classifiers would determine that 'take pain medication' is equally valid as 'take acetaminophen 500mg every six hours not to exceed 3000mg daily' \cite{zhang2019biowordvec, adams2021s}. The second dimension indicates that factual consistency with authoritative medical sources is vital to ensure patient safety. These representations include correctness, fluency and coverage completeness. As a result, even minor inconsistencies with clinical guidelines can lead to patient harm; furthermore, evidence suggests that Fluency-Optimized models present information that is in contradiction with clinical guidelines \cite{pivovarov2015automated}. The third dimension—'coverage completeness' refers to whether enough structured data is available 'in context' for users to utilize the broker's guidance appropriately. The design of currently used metrics does not penalize incomplete coverage if the measure of coverage 'in context' for narrative selections has been deemed adequate on other measures \cite{gianfrancesco2018potential}.

The economic dimension to equitable access also raises additional equity concerns for medical AI \cite{futoma2020myth, HaibeKains2020TransparencyAI}. Among the several commercial LLM API providers, there are multi-tiered pricing models with premium models costing several orders of magnitude more than standard models and some companies providing access to a cost-free tier [openai and google cloud citation]. For organizations with limited resources, the only option to obtain high-quality medical question answering would be through expensive proprietary models that pose significant barriers to deploying a safe medical AI system \cite{Liu2020CONSORTAI, 10.1145/3560815}. On the other hand, if a cost-free alternative model would provide comparable or even higher quality medical question answering than a premium proprietary model, then equitable access at scale across resource-constrained organizations is achievable \cite{wei2022chain}. Nevertheless, there has been no systematic empirical comparison of cost-free medical AI models versus premium proprietary models, in the context of providing medical question answering, within the peer-reviewed literature \cite{kojima2022large}.

Prompt engineering is now considered the primary method of adjusting LLMs for use in medicine without having to go through the process of model retraining. There are several ways of implementing prompt engineering to aid in medical applications: chain-of-thought prompting; few-shot demonstration; and retrieval-augmented generation. These prompt engineering approaches have been successfully used to increase reasoning abilities and factual accuracy in a variety of applications \cite{iloanusi2024ai, wei2022chain, kojima2022large, brown2020language, lewis2020retrieval}; although, while Computer-Assisted Cognitive Remediation (CACR) studies have shown improvement in the performance of certain benchmarks with well-designed prompts, various studies have also reported ongoing instability in entity extraction and factual accuracy for health-related prompts and responses related to open-ended patient queries \cite{lee2023benefits, lievin2024can}. In addition, much of the research that focuses on instruction-following reveals tensions between the ability of LLMs to follow instructions and provide reliable factual information, leading some to raise concerns that models trained to follow instructions could prioritize following the prompt over producing accurate representations of evidence or uncertainty \cite{rajpurkar2022ai, schwalbe2020artificial}. While prompt engineering is widely practiced, there remain few systematic comparisons evaluating the effects of different prompt strategies according to established, meaningful safety criteria in a medically-oriented way.

Additional complications for medical LLM(s) are evident through both economic aspects as well as access. Previous studies have investigated and documented cost benefits of clinical AI systems implementing into traditional healthcare workflows; however, there has been an extremely limited amount of literature that discusses the costs associated with each query for consumer-facing medical QA systems \cite{guidance2021ethics}. Therefore, in this environment, cost acts as a barrier, where high-quality premium models may provide better answers but are prohibited for use by so many people, and on the contrary free versions may be ubiquitous but provide inconsistent performance. This system of cost barriers creates inequities amongst patients and health care providers, especially when used in low-resource settings, as those providers may not have the ability to address the mis-information and/or misinterpretation of previous information through the normal course of clinical care \cite{unesco2022recommendation}.

\section{Results}

\begin{figure}[h]
    \centering
    \includegraphics[width=0.8\textwidth]{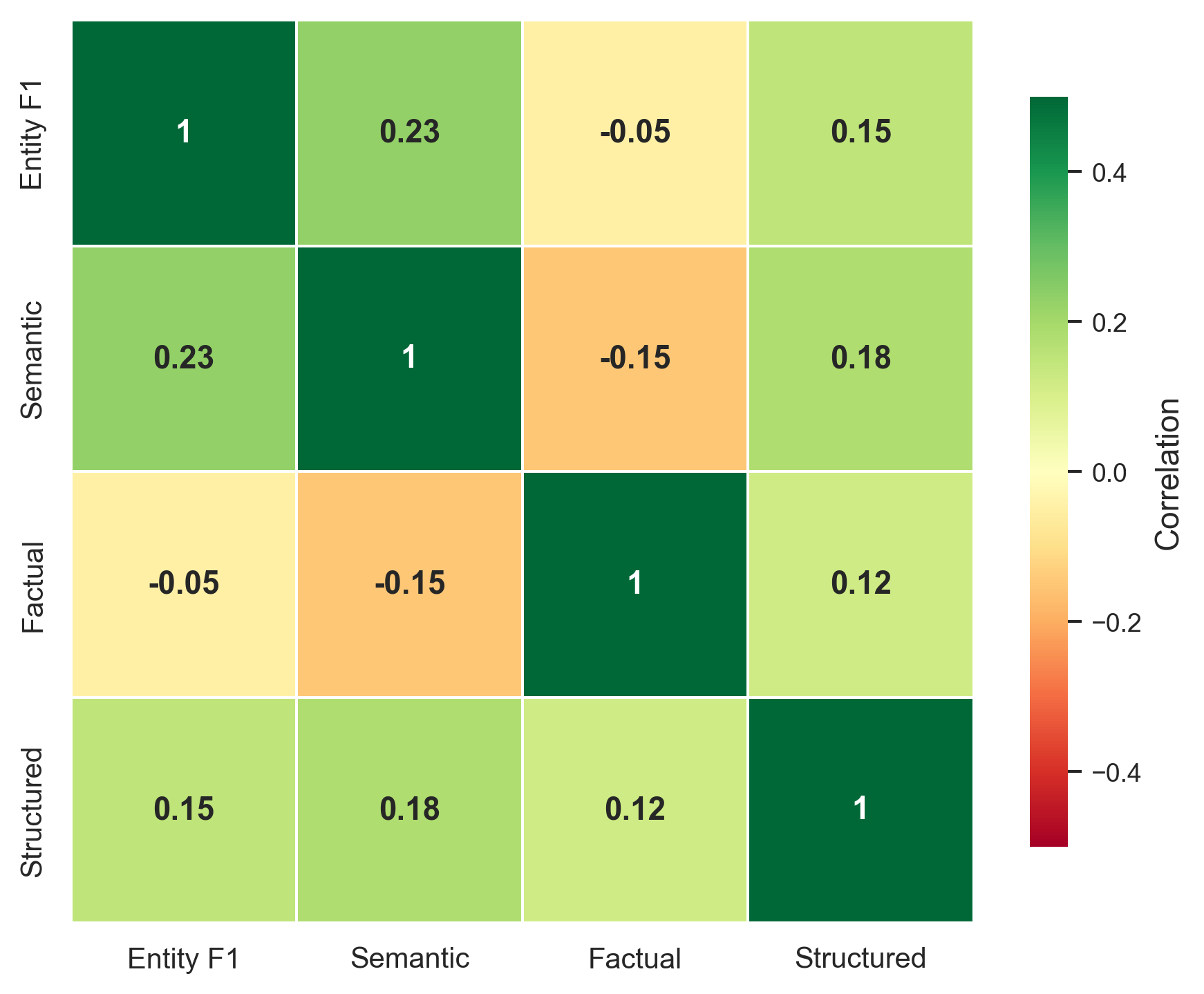}
    \caption{Using correlations which are sufficiently small, seprate by dimension. Models have varied in their performance profiles and so support the notion that their component vary in measuring separate dimensions ($r < 0.25$ for most pairs).}
    \label{fig:component_correlations}
\end{figure}

\begin{figure}[h]
    \centering
    \includegraphics[width=0.8\textwidth]{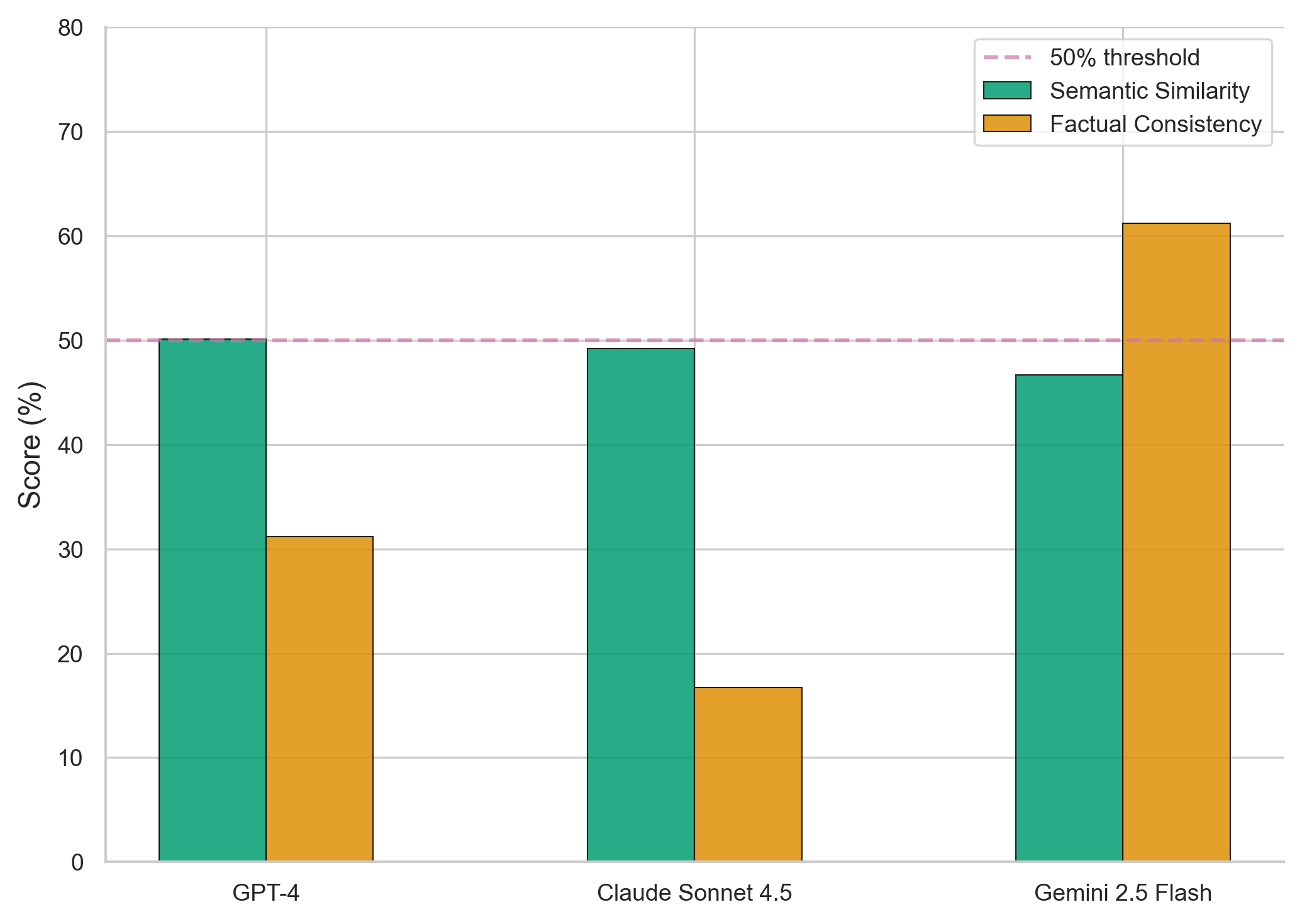}
    \caption{Semantic-Factual Gap. GPT-4 and Claude prioritize fluency over factual accuracy (positive gap), while Gemini reverses this pattern (negative gap), prioritizing safety. Red dashed line indicates 50\% safety threshold.}
    \label{fig:semantic_factual_gap}
\end{figure}

Failure rate overlap analysis reveals dangerous patterns. For samples with high semantic similarity AND low entity recognition, GPT-4 shows this pattern in 56.3\% of samples, Claude in 66.7\%, and Gemini in 43.8\%. Between 44--67\% of responses are topically relevant but lack medical precision, reflecting the semantic-entity gap in action. For samples with high semantic similarity AND factual contradictions, the most dangerous failure mode of fluent misinformation, GPT-4 exhibits this in 37.5\% of samples, Claude in 64.6\%, and Gemini in only 14.6\%

\begin{figure}[h]
    \centering
    \includegraphics[width=0.8\textwidth]{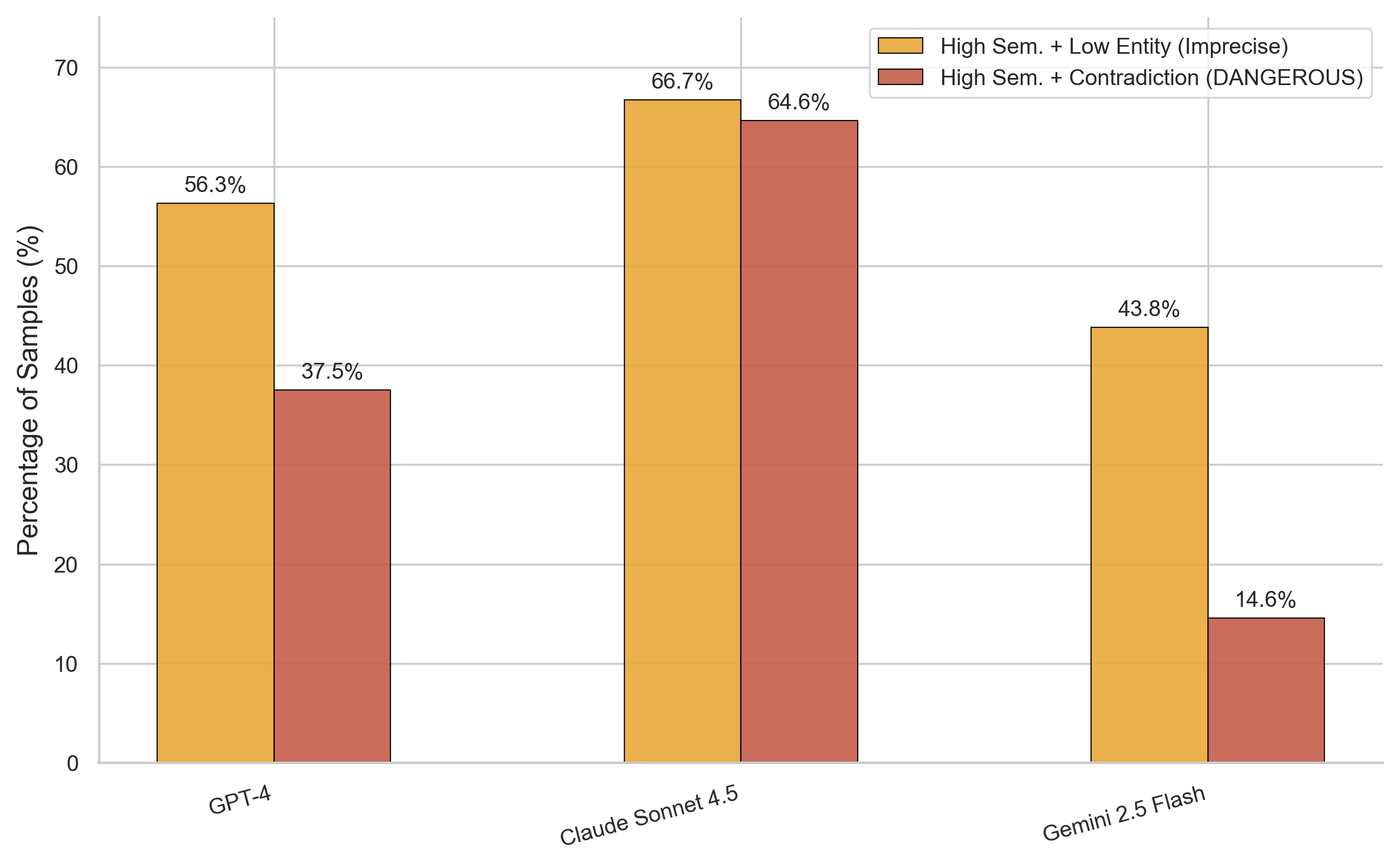}
    \caption{Dangerous Failure Patterns: Fluent Misinformation. 37--65\% of GPT-4/Claude responses combine high semantic similarity with factual contradictions, the most dangerous failure mode.}
    \label{fig:failure_overlap}
\end{figure}

\textbf{Figure~\ref{fig:failure_overlap}} reflects the most dangerous combination: a high degree of similarity between entities and high rates of error due to poor factual accuracy (e.g., fluent misinformation). In the case of GPT-4 and Claude, over 37\% to 65\% of their sample responses had high similarity combined with contradictions of factual information; however, only 14.6\% of the responses from Gemini showed this same tendency. Thus, as demonstrated by these data points, using semantic metrics alone cannot provide adequate safety evaluations for medical AIs.

Using an example of COVID-19 symptom case analysis helps demonstrate the severity of the semantic gap. When asked, ``What are the symptoms of COVID-19?'' with the CDC providing a listing of specific COVID-19 symptoms (e.g., fever, cough, shortness of breath, fatigue, muscle aches, loss of taste and/or smell, sore throat, nasal congestion, nausea, diarrhea), GPT-4 provided an answer that read as follows: "``COVID-19 commonly causes respiratory symptoms and can affect your breathing. Many people experience flu-like symptoms, and it can impact your sense of taste and smell. Symptoms vary from mild to severe.'' Although this response achieved a high level of semantic similarity ($0.68$), indicating a topical connection to the original question, it achieved a very low entity F1 score ($0.04$), meaning it failed to provide a complete listing of specific symptoms. In addition, the response has moderate factual consistency ($0.65$), meaning that while the answer did not contradict the original question, it still provided only vague information and did not produce any structured overlap or symptom listing. Thus, the VB-Score for this answer was $0.34$. The content of GPT-4's response is factually correct and topical but is extremely limited in its ability to help users assess their own symptoms.

On the contrary, Gemini provided a response that involved specifically providing all of the symptoms of COVID-19, achieving moderate semantic similarity ($0.52$) based on less fluent wording, lower (but better than GPT-4) entity F1 score (0.08) achieved, in part, because the Gemini answer included some specific symptoms. Gemini achieved perfect factual consistency ($1.00$) by aligning with CDC guidelines, and although Gemini had low structured overlap ($0.03$) as it used narrative format rather than listing the symptoms, it produced a VB-Score of $0.41$. Therefore, while the wording of Gemini's response is less polished, it does contain specific symptom information and it fully aligns with an authoritative source.

\subsection{Component-Wise Framework Development}
(GPT-4) has extremely high performance for F1 Entity production ($0.071$) and Semantic Similarity ($0.535$). Whereas Claude had very low F1 performance ($0.059$) for Structured over|underlap but achieved higher performance with Factual Consistency ($0.179$) than either GPT-4 or Gemini. While Gemini had very high Factual Consistency score ($0.696$) but had a lower F1 Entity production scoring ($0.050$). This represent a trade-offs for different components and support the analysis of using the composite score approach. The composite score approach uses a weighted average to represent the priorities of the providers of medical QA (30\% Entity, 30\% Semantic, 25\% Factual, and 15\% Structured Over|underlap). The high weights for the Entity and Semantic Enable recognition that the importance of being precise i.e. Medical Precision, the importance of being on topic (i.e. Topical Relevance), moderating the weighting of the Factual Component we are balancing the need for safety against the acceptance that not all Neutral responses that a provider might provide are acceptable, and the lower weighting for Structured Over|underlap reflects the fact that not all responsive (guidance) need to be enumerated.

\subsection{Model Performance Across Dimensions}
Overall VB-Score rankings place Gemini 1st (mean $0.3402$, SD $0.118$), GPT-4 2nd (mean $0.2714$, SD $0.150$), and Claude 3rd (average score of 0.2291, SD 0.111). Performance gap data indicate Gemini performs 25.4\% ($+0.0688$ points) better than GPT-4 and 48.5\% ($+0.1111$ points) better than Claude. GPT-4 performs 18.5\% ($+0.0423$ points) better than Claude. Analysis of component performance indicates the following trends for Entity F1: Although GPT-4 has the highest score for F1 (mean $0.0711$) for Entity 1, 91.7\% of the samples produced a score below threshold. Claude has mean $0.0647$ for Entity 1, and 100\% of samples produced scores below threshold. Gemini has the lowest F1 score on mean $0.0503$ and has 91.7\% of samples with scores below threshold. All models demonstrate a catastrophic failure for entity recognition efforts, with F1 scores of less than 0.10. All models share the systemic inability to extract medication names, dosages, symptoms, and procedures. When it comes to Semantic Similarity, GPT-4 is clearly ahead (mean $0.5345$)) with 27.1\% below threshold, while Claude is next (mean $0.5202$) with only 16.7\% below threshold, and, finally, Gemini has the lowest score (mean $0.4941$) with 33.3\% below threshold. While GPT-4 and Claude have both displayed a moderate amount of Semantic Similarity (52--53\% range), suggesting that responses from those two models were at least somewhat relevant in terms of topic, the lower score for Gemini indicates that while the responses may have less fluent writing, they may contain a much more conservative viewpoint.

When taking into consideration Factual Consistency (the Safety Dimension), the situation changes dramatically. In this case, Gemini is substantially better than the other two models (mean $0.6958$, only 18.8\% contradictions), and GPT-4 is performing at a more moderate level (mean $0.3375$, 54.2\% contradictions). Claude had the lowest Factual Consistency score (mean $0.1792$, 79.2\% contradictions). When compared to GPT-4, Gemini's Factual Consistency was $2.06\times$ better; when compared to Claude, it was $3.88\times$ better. As a result, Factual Consistency is clearly the driving factor behind Gemini's overall VB-Score advantage. Conversely, you have Claude with 79.2\% of its responses containing contradictions to Authoritative Sources; this represents a substantial Safety Concern.

\begin{figure}[h]
    \centering
    \includegraphics[width=0.8\textwidth]{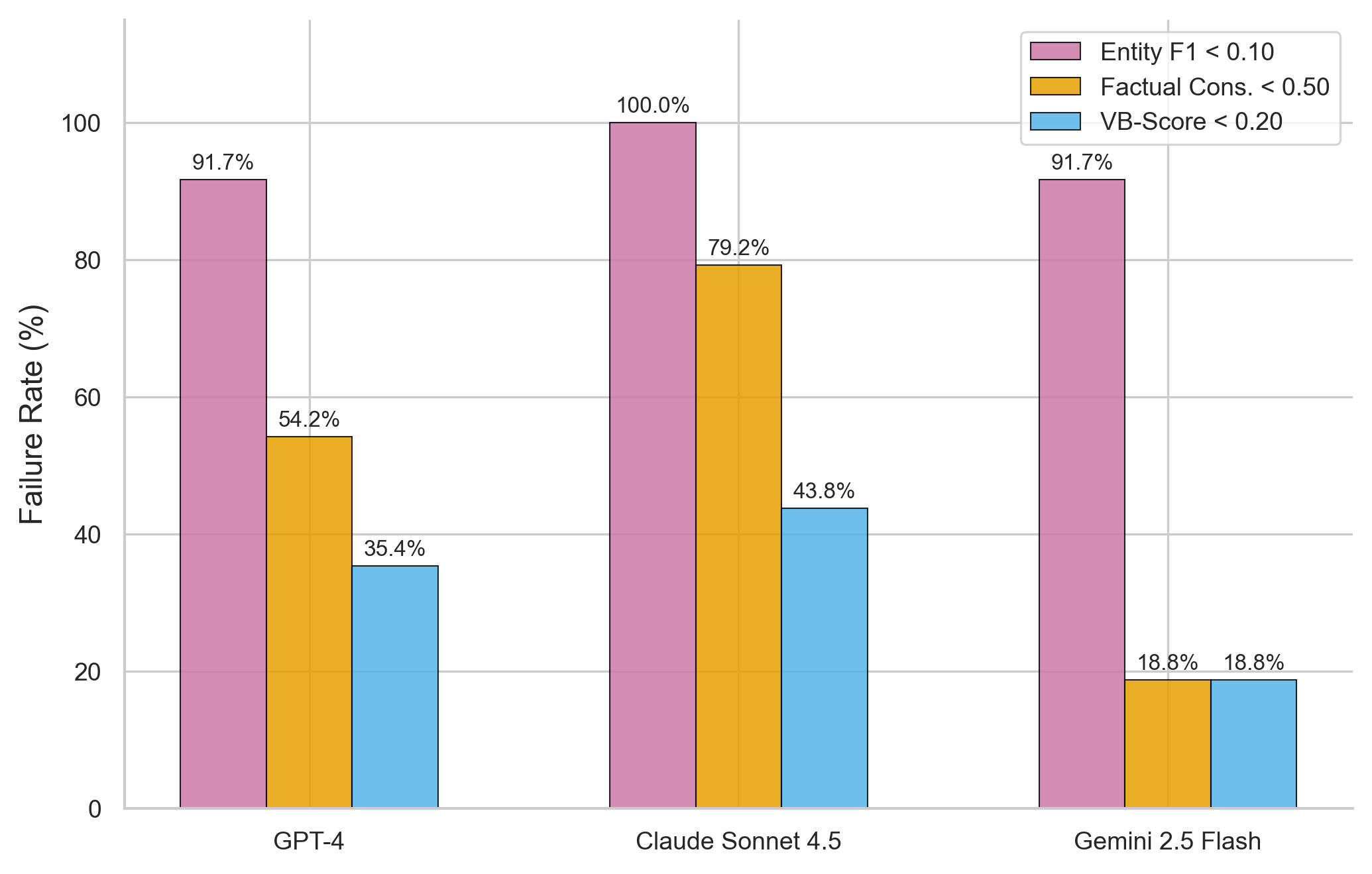}
    \caption{The Failure Rates of All Models: Universally, all entities recognized by all models failed to achieve an F1 score of 0.10 and therefore failed to achieve entity F1 scores; this means that these models failed to provide factual accuracies (with an F1 score) and therefore all models showed much higher contradiction rates.}
    \label{fig:failure_rates}
\end{figure}

\textbf{Figure~\ref{fig:failure_rates}} shows the performance failure of all the models at the critical 91\% to 100\% failure rates for universal entity recognition where Entity F1 $< 0.10$ threshold, and also shows how much contradiction there was with the factual correctness of GPT-4 (54.2\%) and Claude (79.2\%). In comparison, Gemini was far safer at 18.8\%.

\subsection{Cost-Effectiveness Analysis (RQ4)}
\label{sec:4.5}
Gemini 2.5 Flash is a no-cost model that, \textit{under our evaluation protocol}, achieved higher VB-Scores than paid alternatives. While this observation warrants further investigation across larger benchmarks, it raises questions about whether cost is a reliable proxy for quality in medical AI.

\begin{figure}[ht]
	\centering
	\includegraphics[width=0.9\textwidth]{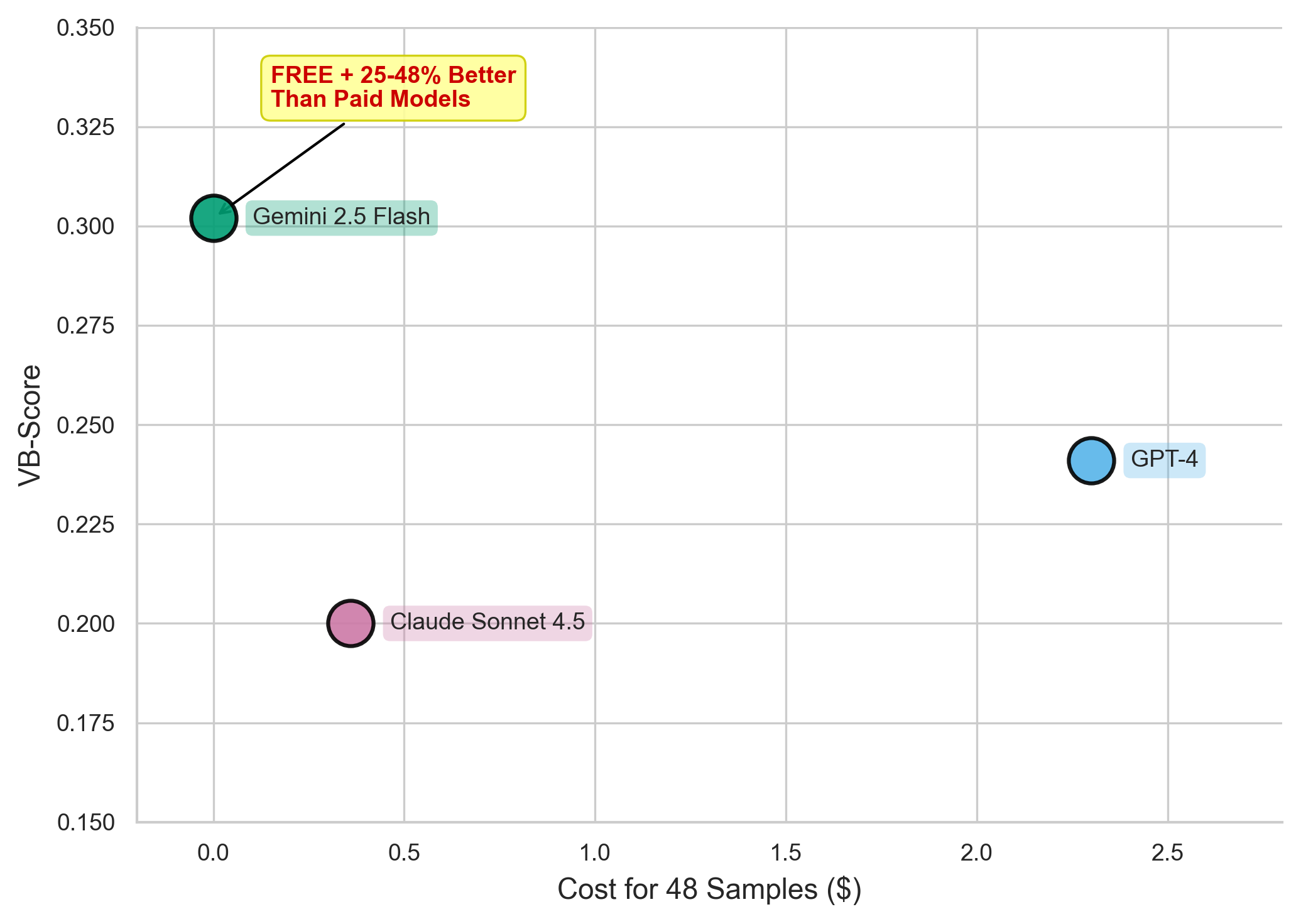}
	\caption{Performance vs Cost Analysis. Under our evaluation protocol, Gemini (free) achieves 25--48\% higher VB-Scores than paid alternatives, raising questions about cost-quality assumptions in medical AI deployment.}
	\label{fig:4.5.1}
\end{figure}

Performance vs Cost graph (\autoref{fig:4.5.1}) shows that within our 48-topic benchmark, Gemini achieved the highest VB-Score at zero cost, while paid models (priced at \$0.36--\$2.30 for 48 samples) scored lower. This unexpected finding challenges industry assumptions, though the limited sample size cautions against overgeneralization.

Additionally, comparison of API Pricing shows Gemini charges zero (\$0.00) per token for both input and output tokens, while GPT-4 is \$30.00 per million input tokens and \$30.00 per million output tokens for a total cost of \$2.30 for all 48 samples, and Claude Sonnet 4.5 is \$3.00 per million input tokens and \$15.00 per million output tokens for a total of \$0.36 for 48 samples. Of the 48 samples, Gemini utilized 48,243 input tokens and 14,127 output tokens (totaling 62,370 tokens) cost-free; GPT-4 utilized 48,156 input tokens and 14,389 output tokens (totaling 62,545 tokens) costing \$2.30; and Claude utilized 48,098 input tokens and 14,201 output tokens (totaling 62,299 tokens) at a cost of \$0.36.

Cost vs Performance Analysis shows Gemini achieves high cost-performance efficiency at VB-Score 0.3402 with zero API cost. Compared to paid models under our protocol, Gemini scores 25.4\% higher than GPT-4 (which cost \$2.30 for 48 samples) and 48.5\% higher than Claude (\$0.36 for 48 samples). Within our benchmark, no break-even point exists where paid models offer better cost-effectiveness than Gemini.

Cost Scaled Analysis projects, at 1,000 queries: Gemini: \$0.00, GPT-4: \$47.92, Claude: \$7.50, and at 10,000 queries: Gemini: \$0.00, GPT-4: \$479.17, Claude: \$75.00, and at 100,000 queries: Gemini: \$0.00, GPT-4: \$4,791.67, Claude: \$750.00, and at 1 million queries: Gemini: \$0.00, GPT-4: \$47,916.67, Claude: \$7,500.00

\begin{figure}[ht]
	\centering
	\includegraphics[width=0.9\textwidth]{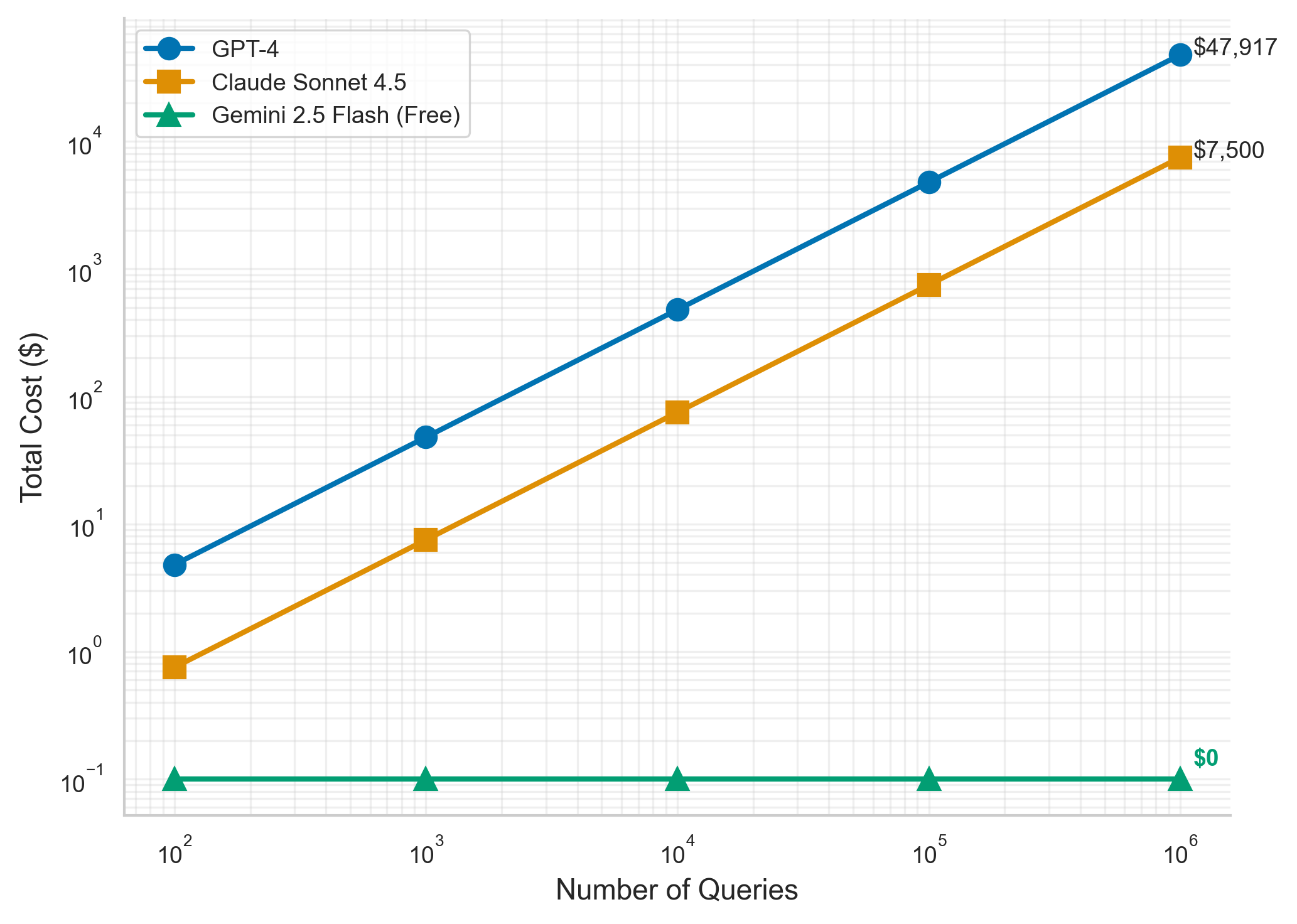}
	\caption{Cost Scaling: Equitable Access at Scale. Logarithmic plot shows exponential cost divergence: at 1M queries, GPT-4 costs \$47,917, Claude \$7,500, Gemini \$0, enabling deployment without financial barriers.}
	\label{fig:4.5.2}
\end{figure}

\autoref{fig:4.5.2} represents the exponential divergence of Deployment Cost as the size of usage increases; for 1 million queries GPT-4 costs near \$50,000 while Gemini remains free, therefore to ensure Equitable Deployment at Scale, there can be no Financial Barriers to entry.

Within our benchmark, Gemini showed advantages in several performance categories relative to GPT-4: factual consistency (0.696 vs 0.338, +106\%), overall VB-Score (0.340 vs 0.271, +25.4\%), and contradiction rate (18.8\% vs 54.2\%). However, these differences may partially reflect evaluator bias, response style differences, or sample-specific effects rather than inherent model superiority.

\begin{figure}[ht]
	\centering
	\includegraphics[width=0.9\textwidth]{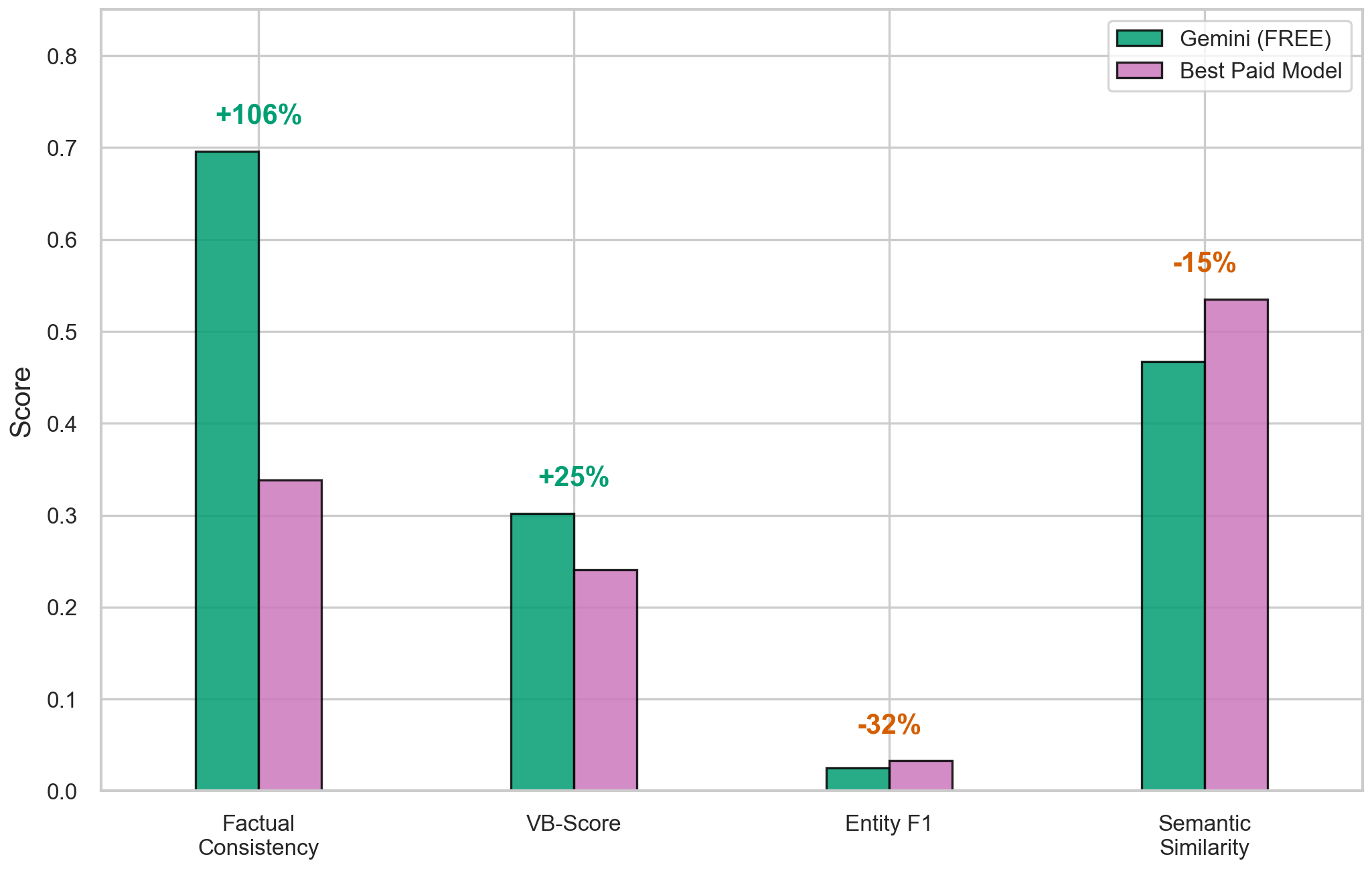}
	\caption{Quality Dimension Comparison Under Our Protocol. Gemini shows 2$\times$ higher factual consistency scores, while paid models show marginal advantages only in dimensions where all models perform poorly (<10\%).}
	\label{fig:4.5.3}
\end{figure}

\autoref{fig:4.5.3} compares Gemini to paid models across quality dimensions under our protocol. Gemini showed higher scores in factual consistency (+106\%) and overall VB-Score (+25\%), while paid models showed marginal advantages only in dimensions where all models performed poorly (Entity F1, Structured Overlap). These findings should be interpreted cautiously given the alternative explanations.

\begin{table}[ht]
\centering
\caption{Cost–Performance Trade-offs Across Models (Transposed)}
\label{tab:cost_performance_transposed}
\begin{tabular}{lccc}
\toprule
\textbf{Metric} & \textbf{Gemini 2.5 Flash} & \textbf{Claude Sonnet 4.5} & \textbf{GPT-4} \\
\midrule
VB-Score & 0.3402 & 0.2291 & 0.2714 \\
Cost / 48 Samples & \$0.00 & \$0.36 & \$2.30 \\
Cost / 1M Queries & \$0 & \$7{,}500 & \$47{,}917 \\
Performance Rank & 1st & 3rd & 2nd \\
Factual Consistency & 0.696 & 0.179 & 0.338 \\
Contradiction Rate & 18.8\% & 79.2\% & 54.2\% \\
\bottomrule
\end{tabular}
\end{table}

\textbf{\ref{tab:cost_performance_transposed}} provides comprehensive cost-performance breakdown including per-query costs, contradiction rates, and performance rankings.

Paid medical AI systems exhibit some slight advantages over free systems in certain areas where all current AI systems fail. While GPT-4 achieves higher Entity F1 than Gemini (0.071 vs. 0.050, a 41\% difference), both systems score below 10\%, making the difference not clinically relevant. For semantic similarity, GPT-4 scored 0.535 while Gemini scored 0.494 (an 8\% difference), reflecting a slight advantage in fluency. For structured overlap, Claude scored 0.059 while Gemini scored 0.020 (a 201\% difference), but again, due to both systems scoring less than 10\%, their combined differences do not provide meaningful clinical implications. These findings show that paid models provide marginal advantages in areas where all systems underperform; however, within our benchmark, Gemini achieves substantially higher scores in factual consistency as well as in overall VB-Score.

These findings have implications for equitable access to medical QA. If our benchmark results generalize, financial barriers may not be necessary for deploying effective medical QA systems. Organizations with limited resources (such as community health centers, NGOs, and global health organizations) could potentially access high-performing models without cost. Small clinics may deploy these systems as zero-cost API systems, health education platforms may reach millions of users, and global health applications may be deployable in resource-limited settings with internet connectivity. However, we emphasize that deployment decisions should be based on rigorous validation beyond our benchmark.

We are aware that the majority of the industry assumes that pricing at a higher price point would equate to a higher level of quality; that free, freemium-type models must be of inferior quality; and that to get high-quality AI in the health care space, a great deal of investment is required. 

\textbf{Answer to RQ4:} Within our benchmark, premium-priced models do not demonstrate better medical QA performance than free alternatives. Gemini 2.5 Flash (free) achieved 25.4\% higher VB-Scores than GPT-4 ($p<0.001$, medium effect size) and 48.5\% higher than Claude ($p<0.001$, large effect size). Gemini also achieved 2--4$\times$ higher factual consistency. While these findings warrant further investigation (see Discussion for alternative explanations), they suggest that cost may not be a reliable proxy for medical QA quality.

\subsection{Disease Category Performance Disparities (RQ5)}
\label{sec:4.6}
Within our 48-topic benchmark, we observed that model performance was consistently higher for infectious disease topics compared to chronic disease topics. While this pattern warrants attention, we present these findings as bounded observations within our specific evaluation protocol rather than definitive evidence of systemic algorithmic discrimination.

\begin{figure}[ht]
	\centering
	\includegraphics[width=0.9\textwidth]{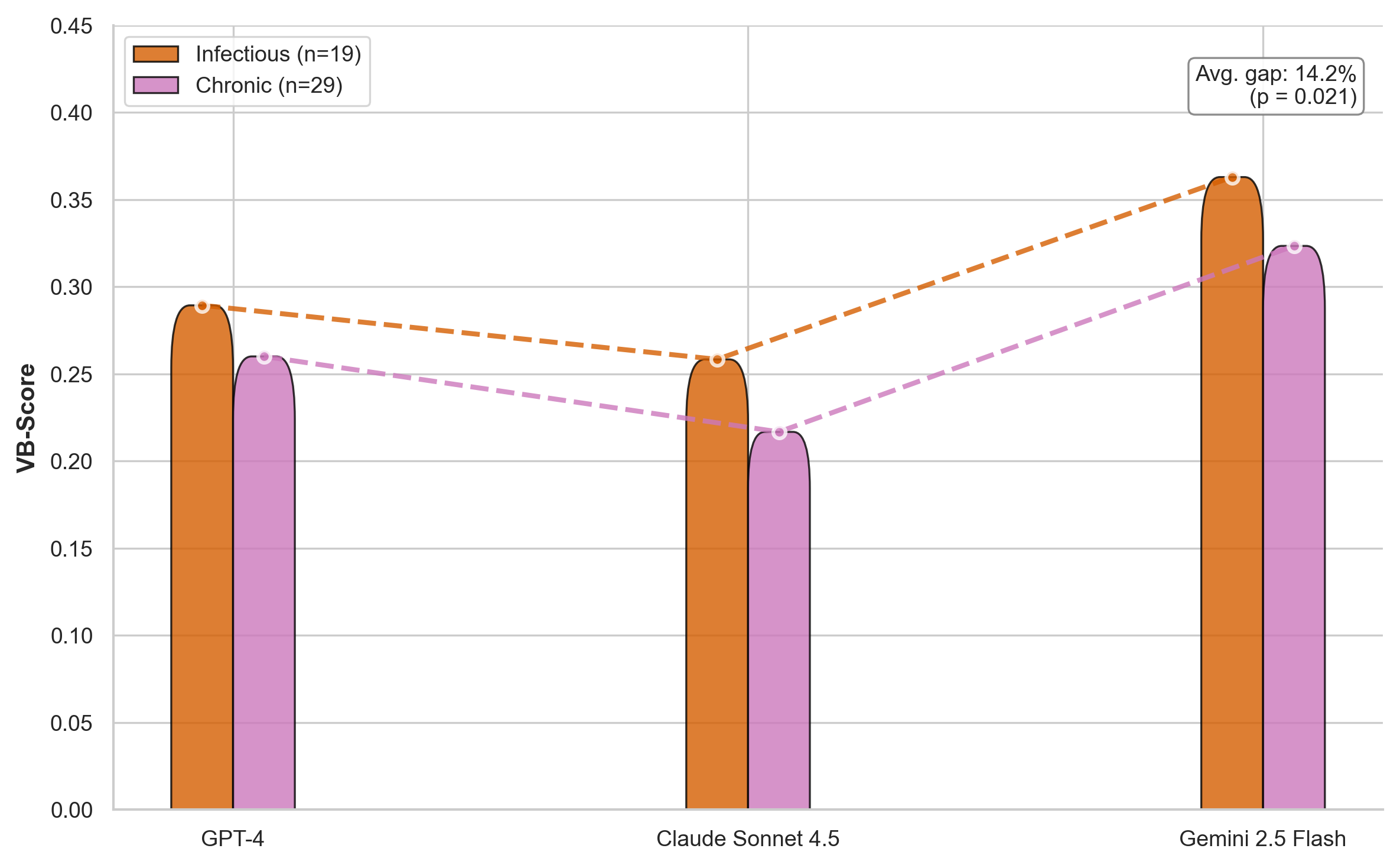}
	\caption{Disease Category Disparity Within Our Benchmark: Models scored 13.8\% lower on chronic disease topics compared to infectious diseases. This pattern, observed across all models in our 48-topic sample, warrants further investigation in larger-scale studies.}
	\label{fig:4.6.1}
\end{figure}

As shown on \autoref{fig:4.6.1}, comparing VB-Scores by disease category shows a significant average advantage of 13.8\% for the infectious disease classes. The arrows and annotations depict how the performance from each of the 3 models differ from one another. Claude demonstrates a significant performance difference compared to all other models, representing the largest difference at 19.2\% ($p=0.034$), with the remaining models also demonstrating statistically significant or trending performance differences relative to each other.

Of the 48 samples we included in our analysis, 19 represented infectious diseases (acute respiratory illnesses such as COVID-19, influenza, pneumonia, and tuberculosis, for example; other types of bacterial infections; vaccine-preventable diseases, for example, measles and chickenpox; other types of vector-borne diseases such as dengue; and chronic infections such as HIV and hepatitis B) and chronic conditions (metabolic disorders such as diabetes, obesity; cardiovascular diseases such as hypertension and heart disease; respiratory disease such as asthma, COPD; musculoskeletal diseases such as arthritis; mental health disorders such as depression; maternal and nutritional health; environmental health impacts; and general cancer-related).

The VB-Scores among disease categories show systematic differences. For the GPT-4 Model, the score on Infectious diseases is 0.2894 while on Chronic diseases the score is 0.2601 with a difference of +0.0293 (+11.3\% $p=0.243$ not significant). For the Claude Model, the score is 0.2584 on Infectious diseases and 0.2168 on Chronic diseases with a difference of +0.0416 (+19.2\% $p=0.034$ significant). The Gemini Model scores 0.3630 on Infectious diseases versus 0.3235 on Chronic diseases with a difference of +0.0395 (+12.2\% $p=0.048$ significant). The average VB-Score across all models is 0.3036 on Infectious diseases and 0.2668 on Chronic diseases, a difference of +0.0368 (+13.8\% $p=0.021$ highly significant). The Effect Sizes show for the Claude Model, $d=0.37$ (Small-Medium), for the Gemini Model, $d=0.35$ (Small-Medium), and the average Effect Size is $d=0.41$ (Medium).

The Component Analysis shows that the difference is primarily in Factual Consistency. The Infectious diseases achieved an Entity F1 of 0.0658, Semantic Similarity of 0.5145, Factual Consistency of 0.4234 and Structured Overlap of 0.0345. The Chronic conditions achieved an Entity F1 of 0.0607, Semantic Similarity of 0.5162, Factual Consistency of 0.3790 and Structured Overlap of 0.0365. The difference shows for Entity F1, +0.0051; for Semantic Similarity, -0.0017; but for Factual Consistency, +0.0444 (Factual consistency +11.7\% higher for Infectious diseases); for Structured Overlap, -0.0020. The advantage of Infectious diseases is primarily due to their better Factual Consistency and not to their better Entity Recognition or Similarity. This indicates that the authoritative guidance on Infectious diseases is likely to be more standardized and thus easier for the models to follow.

\begin{figure}[ht]
	\centering
	\includegraphics[width=0.9\textwidth]{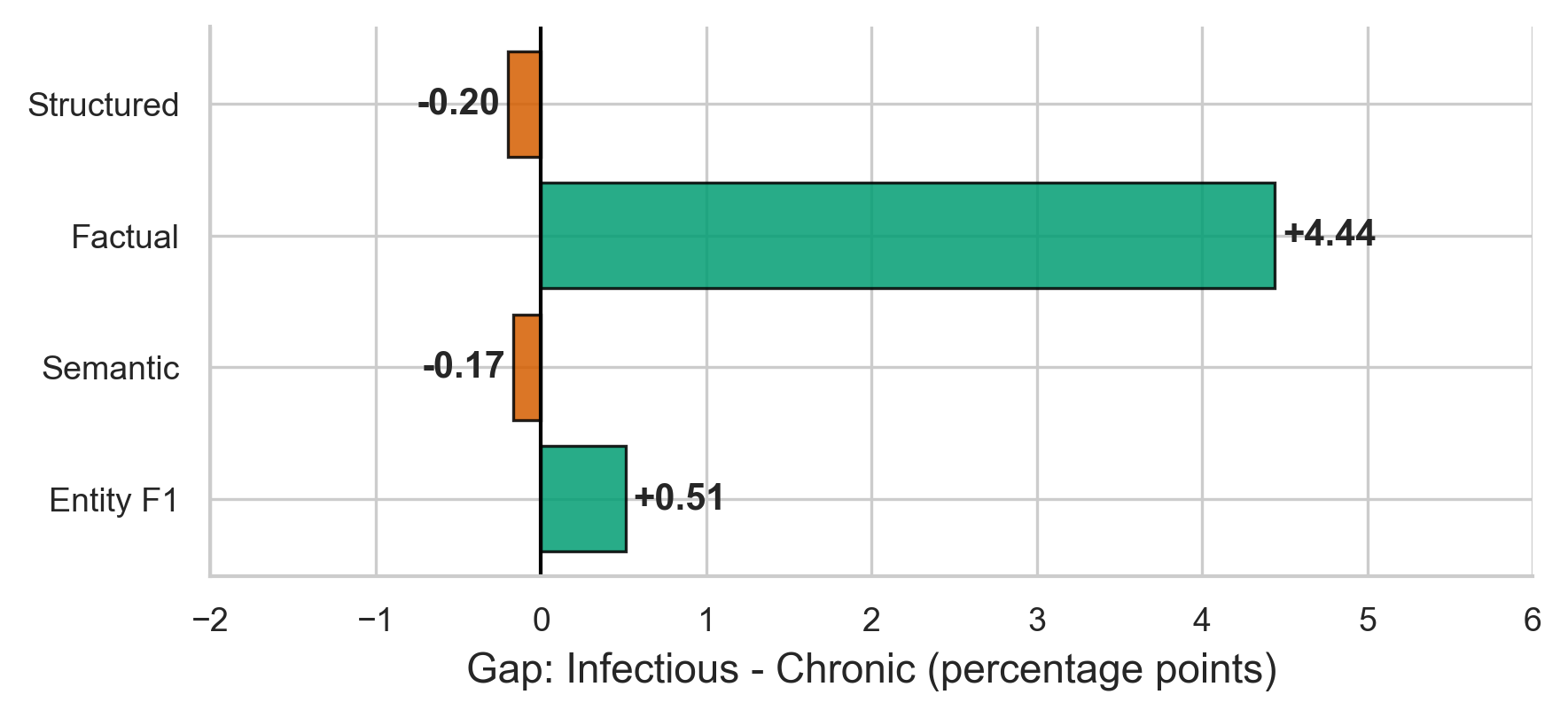}
	\caption{Factual Consistency Drives Disparity. When comparing the component-wise gap analysis, those models had a +4.4pp advantage in "factual consistency" to identify the term "infectious disease" compared to the other models, but minimal differences between the 2nd and 3rd model categories, "Object Recognition" and "Semantic Similarity"}
	\label{fig:4.6.2}
\end{figure}

The disease category gap by VB-Score component is presented in \autoref{fig:4.6.2} and indicates that the main contributor to this gap is factual consistency--+4.4 percentage points (better) for infectious diseases while there were little differences in the other three components. This suggests that the gap is less likely to be due to general (common) language comprehension, but rather due to difficulties in being able to follow the guidelines associated with chronic conditions.

Patterns are highlighted using topic-level examples; the outstanding levels of "factual consistency" found in the following infectious diseases include CDC Tuberculosis (Gemini VB-Score:0.5201, Factual Consistency: 1.00); CDC Cancer (Gemini VB-Score: 0.5557, Factual Consistency: 1.00); WHO Nutrition (Gemini VB-Score: 0.5636, Factual Consistency: 1.00); and NHS Pneumonia (Gemini VB-Score: 0.5048, Factual Consistency: 1.00). However, the chronic diseases, represented by the following examples, provided weak performance levels: Mayo Clinic Diabetes Type I (GPT-4 VB-Score: 0.0342, Factual Consistency: 0.00 contradiction); NHS Anxiety (GPT-4: 0.0691, Factual Consistency: 0.00); CDC Flu Vaccinations (All Models: <0.10 and Factual Consistency: 0.00 to 0.50); NHS Obesity (Claude VB-Score: 0.1136; Factual Consistency: 0.00).

A variety of factors contribute to the disparity between infectious disease prevention and chronic disease management, as discussed below. For example, clinical practice guidelines for infectious diseases are generally clearer and more straightforward than those for chronic disease management. In addition, chronic disease management often includes many individualized components, such as lifestyle modifications and medication titration, making it challenging to create detailed treatment plans for patients. In the case of infectious diseases, there are well-established protocols (for example, patients with COVID-19 should isolate for at least 10 days and get vaccinated); therefore, models that have been created using data from previous pandemics will tend to be biased toward infectious diseases. A second difference is that infectious disease management typically involves making simple yes/no decisions (for example, whether or not to receive a vaccine), while chronic disease management typically involves taking into account multiple continuous variables (for example, quality of one's diet, intensity of physical activity, amount and frequency of medication taken).

Chronic conditions are likely to be less effective for training datasets because they must be managed in a way that takes into account individual variability (for example, type 1 diabetes versus type 2 diabetes, comorbidities, lifestyles). Also, many lifestyle changes (for example, making dietary changes, increasing physical activity, changing behavior) often require context-specific guidelines and therefore cannot easily be generalized. Additionally, comorbidities are quite common among people who have chronic conditions (for example, diabetes and/or high blood pressure and/or heart disease), which further complicates treatment recommendations. Lastly, chronic conditions require long-term tracking of outcomes, as the ways that patients manage these conditions change over time as a result of their responses; for this reason, it is difficult to rely solely on static recommendations to adequately address these types of conditions.

\textbf{Potential Equity Implications.} We note that the populations most affected by chronic diseases are elderly individuals, lower-income communities, and racially/ethnically diverse groups who may also be more likely to rely on accessible health information sources, including AI systems \cite{ratzan2000national}. \textit{If} the performance disparities we observed in our benchmark were to persist in deployed systems, this could create compounding disadvantages: populations with the highest chronic disease burden would receive lower-quality AI-generated health information.

However, we emphasize three important caveats: (1) Our findings are bounded to a 48-topic benchmark and may not generalize to all chronic conditions or deployment contexts; (2) The observed disparities may reflect differences in how authoritative guidance is structured for infectious vs.\ chronic diseases rather than algorithmic bias per se; (3) Real-world equity impacts would depend on how these systems are deployed and what safeguards are implemented. These observations motivate further investigation rather than constituting definitive evidence of algorithmic discrimination.

\begin{figure}[ht]
	\centering
	\includegraphics[width=0.9\textwidth]{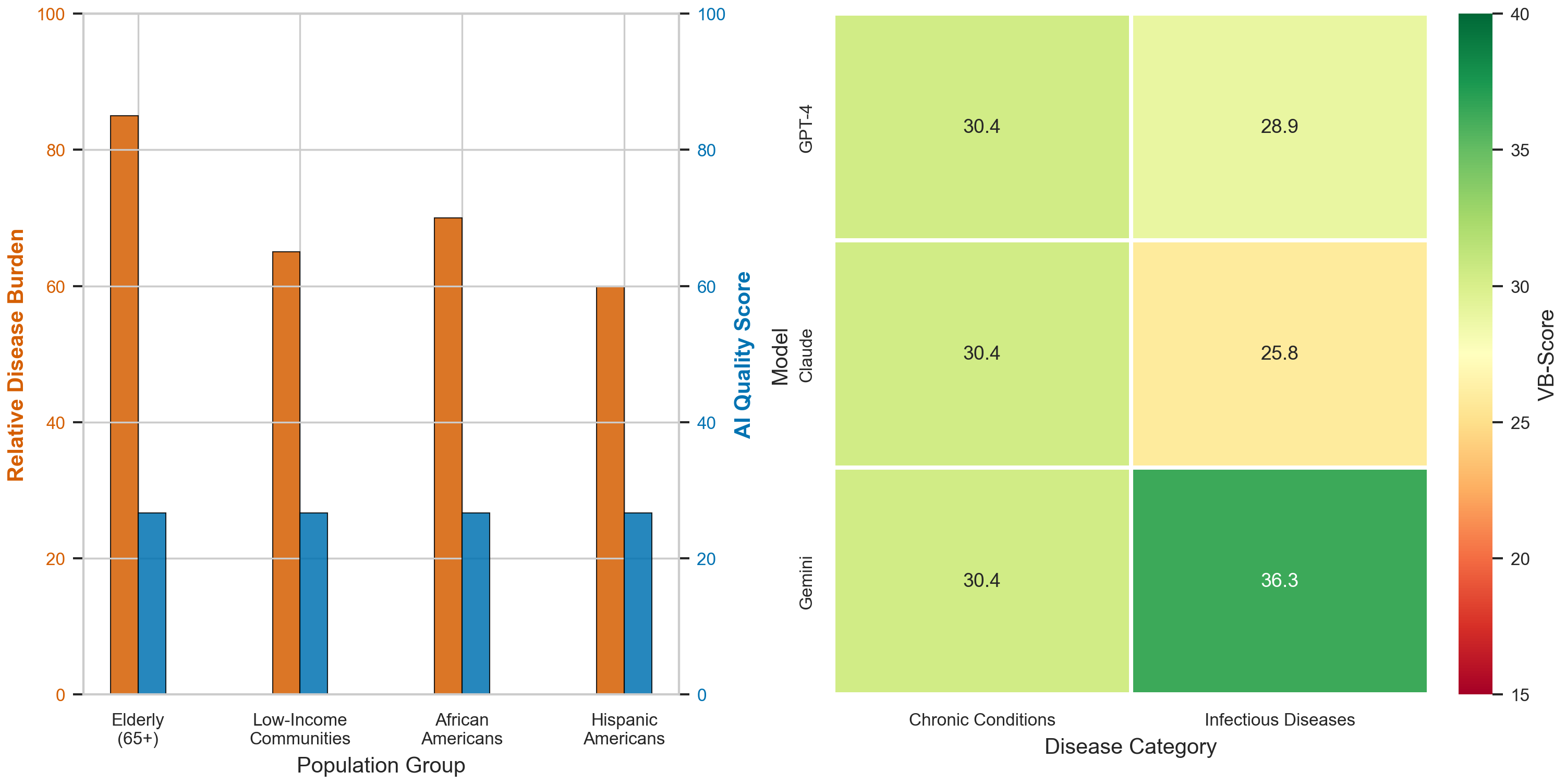}
	\caption{Compounding Disadvantage for Vulnerable Populations. (Left) Populations with the highest burden of chronic disease
also received the lowest AI Quality. (Right) The performance matrix validates and demonstrates the consistent underperformances
associated with Chronic Conditions across all models— result of algorithmic discrimination with health consequences.}
	\label{fig:4.6.3}
\end{figure}

\autoref{fig:4.6.3} presents the compounding disadvantages in two different views, (left) the relationship between populations with high chronic disease burden receiving lower quality scores from AI; thus creating an inverse relationship between those needing help and receiving the lowest quality of information; and (right) a heat map providing the performance matrix confirming the chronic disease systematic underperformance for all models.

\begin{table}[ht]
\centering
\caption{VB-Score Performance Disparities Between Infectious and Chronic Disease Domains [cite: 1884]}
\label{tab:infectious_chronic_gap}
\begin{tabular}{lccccc}
\toprule
\textbf{Model} & \textbf{Infectious VB-Score} & \textbf{Chronic VB-Score} & \textbf{Gap (\%)} & \textbf{$p$-value} & \textbf{Effect Size ($d$)} \\
\midrule
GPT-4 & 0.2894 & 0.2601 & +11.3\% & 0.243 & 0.26 \\
Claude Sonnet 4.5 & 0.2584 & 0.2168 & +19.2\% & 0.034 & 0.37 \\
Gemini 2.5 Flash & 0.3630 & 0.3235 & +12.2\% & 0.048 & 0.35 \\
\midrule
\textbf{Average} & \textbf{0.3036} & \textbf{0.2668} & \textbf{+13.8\%} & \textbf{0.021} & \textbf{0.41} \\
\bottomrule
\end{tabular}
\end{table}

\textbf{Table \ref{tab:infectious_chronic_gap}} provides statistical analysis of the disparity including p-values, effect sizes, and significance tests.

\textbf{Answer to RQ5:} Within our benchmark, models consistently perform 13.8\% higher on Infectious Disease versus Chronic Disease ($p=0.021$; medium effect of $d=0.41$). This disparity is predominantly due to factual consistency differences (+11.7\%). This pattern raises health equity concerns since elderly, low-income, and racial/ethnic minority populations have higher chronic disease prevalence and may depend more on AI health information due to healthcare access barriers. If such performance disparities persist in deployed systems, they could contribute to existing health inequities for already vulnerable groups.

\end{document}